\documentclass[twocolumn,aps,prc,superscriptaddress]{revtex4-2}
\usepackage{color,amsmath,amssymb,epsfig,graphicx}
\usepackage{bm}
\usepackage[colorlinks=true,linkcolor=blue,filecolor=blue,urlcolor=blue,citecolor=blue]{hyperref}
\usepackage{mathptmx, courier, pifont}
\usepackage[scaled=0.92]{helvet}
\usepackage[T1]{fontenc}
\usepackage{textcomp}
\usepackage{multirow}
\usepackage{longtable}

  \def\ack{\,|\,}
\begin{document}
\title{Triaxial projected shell model approach for negative parity states in even-even nuclei}
\author{ Nazira Nazir}
\affiliation{Department of Physics, University of Kashmir, Srinagar, 190 006, India}
\author{S.~Jehangir} \email{sheikhahmad.phy@gmail.com}
\affiliation{ Department of  Physics, Islamic University of  Science and Technology, Awantipora, 192 122, India}
\author{ S.P. Rouoof}
\affiliation{ Department of  Physics, Islamic University of  Science and Technology, Awantipora, 192 122, India}
\author{ G.H.~Bhat}
\affiliation{Department of Physics, SP College  Srinagar, Jammu and Kashmir, 190 001, India}
\affiliation{Cluster University Srinagar, Jammu and Kashmir, Srinagar, Goji Bagh, 190 008, India}
\author{ J.A. Sheikh}
\affiliation{Department of Physics, University of Kashmir, Srinagar, 190 006, India}
\affiliation{ Department of  Physics, Islamic University of  Science and Technology, Awantipora, 192 122, India}
\author{ N. Rather}
\affiliation{ Department of  Physics, Islamic University of  Science and Technology, Awantipora, 192 122, India}
\author{Manzoor A. Malik}
  \affiliation{Department of Physics, University of Kashmir, Srinagar, 190 006, India}
  \affiliation{ Department of  Physics, Islamic University of  Science and Technology, Awantipora, 192 122, India}

\date{\today}

\begin{abstract}

  The triaxial projected shell model (TPSM) approach is generalized to investigate the negative parity band structures in
  even-even systems. In the earlier version of the TPSM approach, the quasiparticle excitations were restricted to one major oscillator
  shell and it was possible to study only positive parity states in even-even systems. In the present extension, the excited
  quasiparticles are allowed to occupy two major oscillator shells, which makes it possible to generate the negative parity states.
  As a  major application of this  development, the extended approach is applied to elucidate the negative parity high-spin band structures
  in $^{102-112}$Ru and it is shown that energies obtained  with neutron excitation are slightly lower than the energies calculated
  with proton excitation. However, the calculated aligned angular momentum ($i_x$) clearly separates the
  two spectra with neutron $i_x$ in reasonable agreement with the empirically evaluated $i_x$ from the
  experimental data, whereas proton $i_x$ shows large deviations. Furthermore, we have also deduced the
  transition quadrupole moments from the TPSM wavefunctions along the negative-parity yrast- and yrare- bands
  and it is shown that these quantities exhibit rapid changes in the bandcrossing region.

\end{abstract}
\maketitle

\section{Introduction}

To characterize the rich band structures observed in atomic nuclei is one of the
main research themes in nuclear structure physics \cite{BMII,sf01}. Major advancements in the experimental techniques have made it feasible to populate multiple high-spin band structures, and in some nuclei more than fifty band structures have been reported \cite{simpson23,ollier11}. The description of this wealth of nuclear structure information is a major challenge to nuclear structure models \cite{SGH16}. In recent years, tremendous progress has been made in the spherical shell model (SSM) description of the nuclear properties \cite{otuska20,Brown22,Poves12}. It is now possible to apply SSM approach to medium mass nuclei, but studying high-spin band structures in heavy-mass region is still  beyond the scope of this microscopic model. To describe the high-spin band structures, it is imperative to include, atleast, two-major oscillator shells as aligning particles occupy the high-j intruder orbitals. To perform the SSM calculations with two-major oscillator shells
for heavier nuclei is beyond the reach of computational resources presently available \cite{langanke2012}.

On the other hand, although several major oscillator shells are considered in density functional approaches, but most of the calculations are restricted to investigate the ground-state properties only \cite{JAS21}. In order to study the high-spin band structures, the angular-momentum projection is required to be performed from the intrinsic mean-field state \cite{ring80}. However, this approach is plagued with the singularity problem due to the reason that most of the modern energy density functionals are fitted to the experimental data
with fractional density
dependence and employ different forces in particle-hole and particle-particle channels \cite{Bender09,Duguet09,dobacz07}. It is to be added that in some recent works \cite{VRETENAR05,meng,wang2023}, the angular-momentum projection has also been performed in density functional theory (DFT) with projection after variation and the singularity problem does not appear to show up in these studies. However, it has been discussed in Ref.~\cite{JAS21} that projected results will contain spurious components that need to be examined. 

Considering the above problems associated with the SSM and DFT approaches, triaxial projected shell model (TPSM)  approach has become a tool  of choice to investigate the high-spin band structures in well deformed and transitional nuclei \cite{JS99,nazira22,jeh21,ishfaq20,jehangir12}. The advantage of this approach is that computational resources involved are quite modest and it is possible to perform a systematic study of a large set of atomic nuclei. As a matter of fact, several systematic investigations have been performed for chiral, wobbling and $\gamma$-vibrational band structures observed in triaxial nuclei \cite{SGH16,nazira23,nazira22,jeh21,ishfaq20,jehangir12,SGH16,JG12}. The model space in the TPSM approach is spanned by  multi-quasiparticle basis states which allows to investigate high-spin band structures. In the original version of the TPSM approach, the model space was quite limited \cite{JS99} but in recent applications \cite{jehangir12,jeh21,nazira22,nazira23,JS21}, we have generalized the basis space to include higher order quasiparticle states. For instance, for even-even systems \cite{jehangir12}, the TPSM approach has been generalized to include four-neutron and four-proton quasiparticle basis states. This extension allows to investigate the high-spin properties in even-even systems beyond the second band crossing.

Nevertheless, in all the extended versions of the TPSM approach the basis configurations are constructed from one major oscillator shell only, although the vacuum configuration is generated from all the three major shells considered in the model.
The justification is that the aligning particles occupy high-j intruder subshell and in order to describe band crossing, it is sufficient to consider quasiparticle excitations only from one major shell containing the intruder orbital.

For even-even systems, the restriction of the quasiparticle excitations from one oscillator shell gives rise to only positive parity states and in order to generate the negative parity states, the quasiparticle excitations need to be considered from two oscillator shells having different parities for the single-particle states. The purpose of the present work is to develop the generalized TPSM approach with the quasiparticle excitations from two major oscillator shells. There is a considerable data available for negative parity bands in even-even systems \cite{NPRU,NPRU2,NPRU3,NP4,NP5,NP6,Che04,Jiang03,dejb95}. However, there have been very limited theoretical calculations to investigate these band structures.

In the present work, we shall focus on the application of the new development to neutron-rich nuclei around A $ \sim $ 110. The properties of these nuclei are studied by measuring prompt $\gamma$-rays emitted by secondary fragments produced by spontaneous and induced fission of $^{252}$Cf source \cite{NPRU,ZHU07}. The weak transitions in the excited negative parity bands are identified through triple and higher order coincidence techniques \cite{SNY13} using the state-of-the-art detector arrays. The ground-state positive parity bands in this region are known to have strong prolate shapes \cite{Fotiades1997}, and the negative parity doublet bands identified in some nuclei are proposed to originate from chiral symmetry breaking mechanism \cite{NPRU}. In our earlier publications, we have  studied the positive parity bands in this mass region \cite{JS21,nazira23,Chanli15} using the TPSM approach and in the present work, we shall focus on the negative parity band structures. Some preliminary results of the present approach for the observed negative parity band structures in $^{106,108}$Mo were published with the experimental group \cite{Musangu18}. However, in this work only neutron-excitation were considered. In the present work, both neutron- and proton-excitations are included in the model space.  The remaining manuscript is organized in the following manner. In the next section, we provide a few details of the extended TPSM approach for negative parity bands and some explicit expressions of the matrix elements are included in the appendix. In section III, the results obtained for Ru-isotopes are presented and discussed, and finally the present work is summarized in section IV.

\begin{table}
\caption{ Axial and triaxial quadrupole deformation parameters
$\epsilon$ and $\gamma = \textrm{tan}^{-1}{(\epsilon'/\epsilon)}$  employed in the TPSM calculation. Axial 
deformations $\epsilon$ have been considered from \cite{moller08} with some adjustment as discussed in the text. The nonaxial values ($\gamma$) 
are chosen in such a way that observed data is reproduced. }
\resizebox{1\columnwidth}{!}
{
\tiny
  \begin{tabular}{ccccccc}
  \hline\hline
  
  &$^{102}$Ru & $^{104}$Ru  & $^{106}$Ru  & $^{108}$Ru  & $^{110}$Ru  & $^{112}$Ru\\
\hline $\epsilon$ &{0.220}     & 0.270       & 0.280      &  0.275     &  0.275    &0.270 \\
      $\gamma$    &30$^{\circ}$& 26 $^{\circ}$&  25$^{\circ}$& 26$^{\circ}$& 30$^{\circ}$& 30$^{\circ}$\\\hline\hline
\end{tabular}\label{tab1}
}
\end{table}

\section{Triaxial Projected Shell Model Approach}
TPSM approach is similar to the SSM technique with the difference that deformed basis are used instead of the spherical one's \cite{JS99,KY95}.
The deformed basis are the optimum basis states to study deformed nuclei, and in the TPSM approach these are generated by solving the three-dimensional Nilsson
mean-field potential \cite{nilsson95}. The pairing interaction is then considered in the Bardeen-Cooper-Scrieffer (BCS)  approximation \cite{ring80}.
The Nilsson + BCS wavefunctions thus constructed form the basis configuration in the TPSM approach. The vacuum state is then constructed by considering valence neutrons and protons to occupy three major oscillator shells \cite{JS99}. However, the quasiparticle excitations are considered from one major oscillator shell only. For instance, to investigate the high-spin band structures in mass $\sim$ 110 region, quasiproton (quasineutron) excitations are considered from the N = 4 (5) shells which contain the $1g_{9/2}(1h_{11/2})$ shell that is responsible for proton (neutron) alignments in this region. This restriction allows to study only positive parity band structures in even-even systems and in order to describe the negative parity band structures, valence pair of particles need to be placed in two different oscillator shells having opposite parities.

In the present work, we have generalized the TPSM approach with valence neutrons and protons occupying  different shells. The extended basis space is composed of:
\begin{figure}[htb]
 \centerline{\includegraphics[trim=0cm 0cm 0cm
0cm,width=0.55\textwidth,clip]{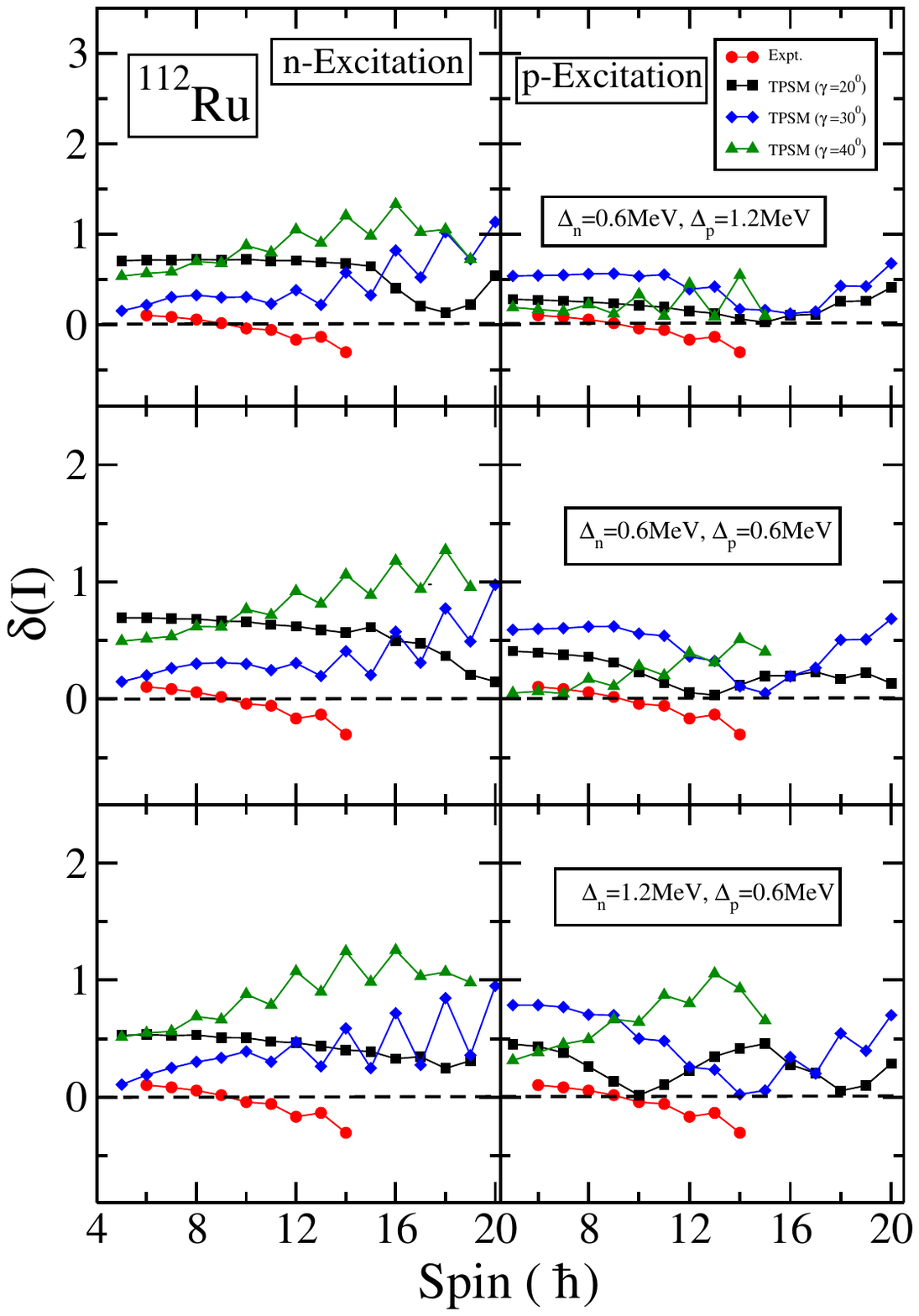}} \caption{(Color
   online) Energy difference between negative-parity yrast- and yrare- bands for same spin, $I$, $\delta(I)=(E_2(I)-E_1(I))$ in $^{112}$Ru.
 }
\label{del112}
\end{figure}
\begin{figure}[htb]
 \centerline{\includegraphics[trim=0cm 0cm 0cm
0cm,width=0.55\textwidth,clip]{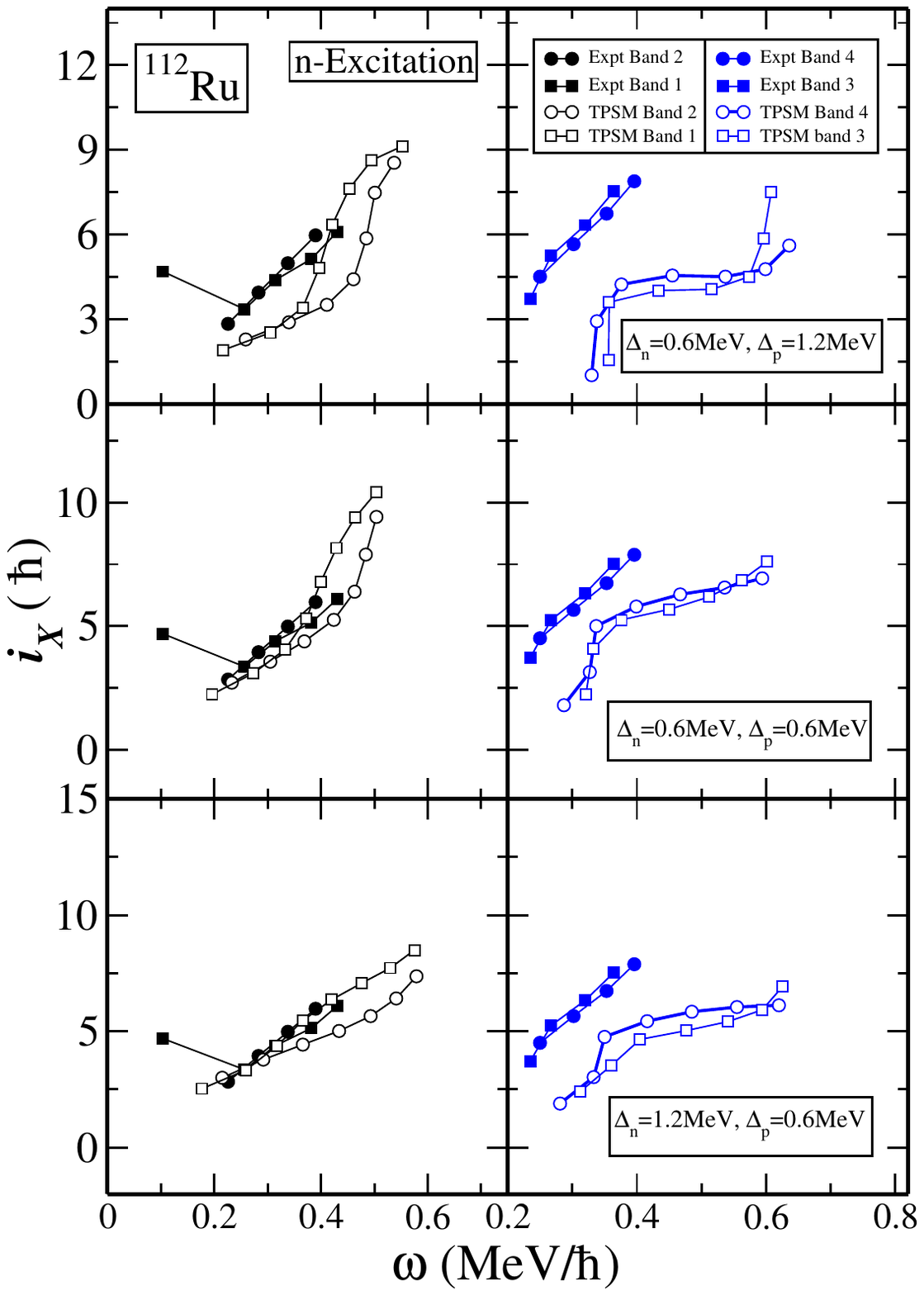}} \caption{(Color
   online) Comparison of the aligned angular momenta,
  $i_X =i_x(\omega)-i_{x,ref}(\omega)$, where $\hbar \omega = \frac{E_{\gamma}}{I^i_{x}(\omega)-I^f_{x}\omega}$,$I_x\omega = \sqrt{I(I+1)-K^2}$ and $i_{x,ref}(\omega)=\omega(J_0 +\omega^2 J_1)$. The reference band Harris parameters
used are $J_0 = 14$ and $J_1 = 15$, obtained from the measured energy
levels as well as those calculated from the TPSM results
for neutron excitation in  $^{112}$Ru.
   }
\label{alignneut112}
\end{figure}
\begin{eqnarray}
  \hat P^I_{MK}~a^\dagger_{n_{1}} a^\dagger_{n_{2}^\prime} |\Phi\rangle;\nonumber \\ 
\hat P^I_{MK}~a^\dagger_{n_{1}} a^\dagger_{n_2^\prime} a^\dagger_{n_3^\prime} a^\dagger_{n_4^\prime} |\Phi\rangle;\nonumber\\
\hat P^I_{MK}~a^\dagger_{n_{1}} a^\dagger_{n_2^\prime} a^\dagger_{p_1^\prime} a^\dagger_{p_2^{\prime}} |\Phi\rangle; \nonumber\\
\hat P^I_{MK}~a^\dagger_{p_1} a^\dagger_{p_{2}^\prime} |\Phi\rangle;\nonumber \\
\hat P^I_{MK}~a^\dagger_{p_{1}} a^\dagger_{p_{2}^\prime} a^\dagger_{p_3^\prime} a^\dagger_{p_4^\prime} |\Phi\rangle;\nonumber\\ 
 \hat P^I_{MK}~a^\dagger_{p_1} a^\dagger_{p_2^\prime} a^\dagger_{n_1^\prime} a^\dagger_{n_2^\prime} |\Phi\rangle,\label{basis}
\end{eqnarray}
\begin{figure}[htb]
\centerline{\includegraphics[trim=0cm 0cm 0cm
0cm,width=0.55\textwidth,clip]{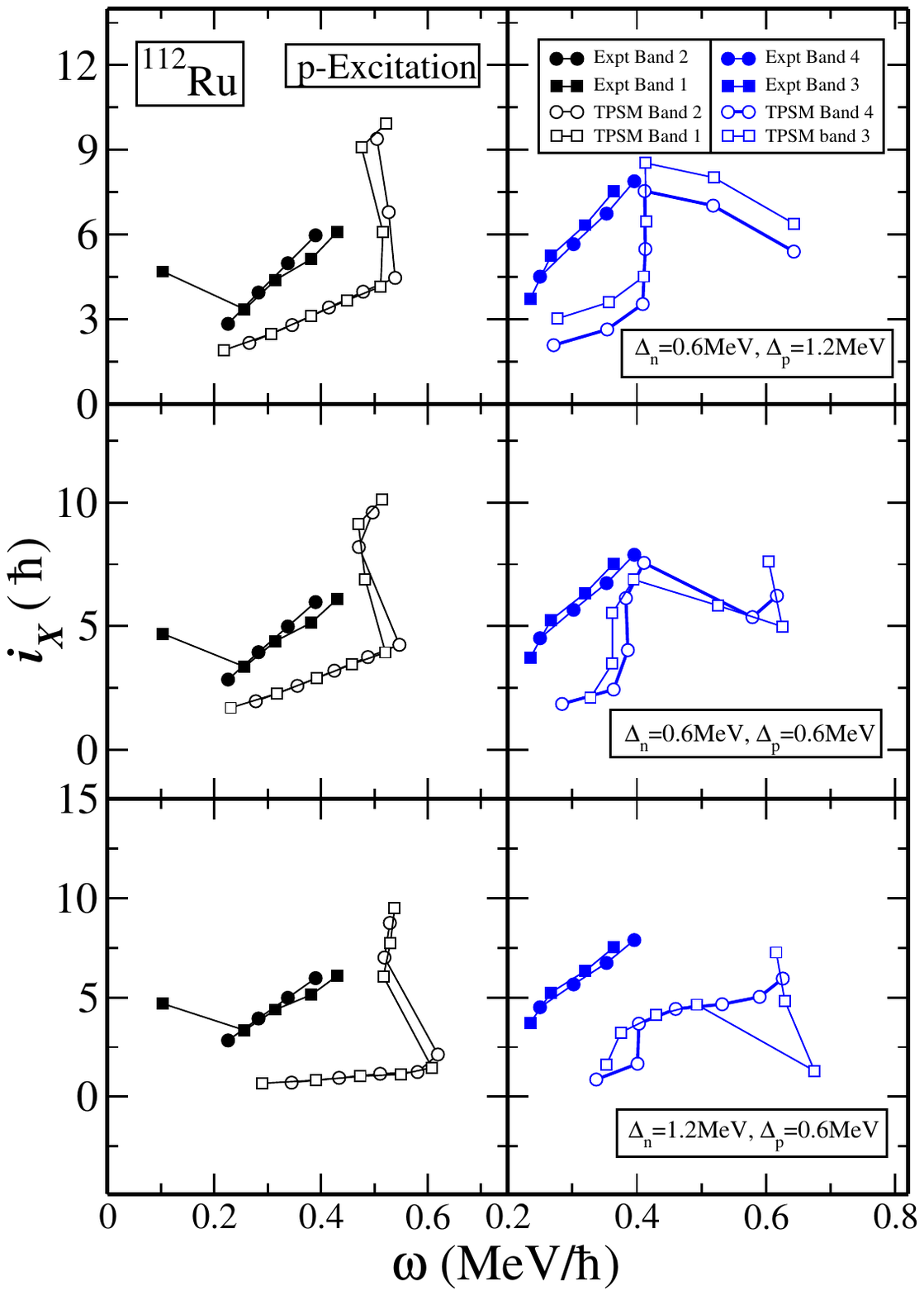}} \caption{(Color
   online) Comparison of the aligned angular momenta,
   $i_X =i_x(\omega)-i_{x,ref}(\omega)$, where $\hbar \omega = \frac{E_{\gamma}}{I^i_{x}(\omega)-I^f_{x}\omega}$,$I_x\omega = \sqrt{I(I+1)-K^2}$ and $i_{x,ref}(\omega)=\omega(J_0 +\omega^2 J_1)$. The reference band Harris parameters
used are $J_0 = 14$ and $J_1 = 15$, obtained from the measured energy
levels as well as those calculated from the TPSM results
for proton excitation in  $^{112}$Ru.
  }
\label{alignprot112}
\end{figure}
where the neutron (proton) major oscillator shells employed are
designated by the quantum numbers $n$ ($p$) with neutrons (protons)
occupying two different oscillator shells, $n (p)$ and $n^\prime (p^\prime)$. We have considered excitations in both proton and
neutron sectors, however, the separable interaction employed doesn't mix these excitations and the two can be diagonalized separately. It is
demonstrated in the appendix that matrix elements between neutron and proton excitations vanish.

It has been demonstrated that two-neutron and two-proton excitations are almost at the same energy \cite{NPRU2}, it is therefore necessary to consider both neutron and proton excitations in the TPSM basis. Further, two-neutron and two-proton aligning configurations have been added as negative parity bands in some nuclei that  have been populated up to quite high-spin and the band crossing phenomena have been observed. $\ack\Phi\rangle$ in Eq.~(\ref{basis}) is the quasiparticle vacuum state which has positive parity, and the three-dimensional angular-momentum projection operator, $\hat P^I_{MK}$, is given by \cite{ring80,HS79,HS80}
\begin{equation}
\hat P^I_{MK} = {2I+1 \over 8\pi^2} \int~d\Omega\,
D^{I}_{MK}(\Omega)\, \hat R(\Omega),
\label{PD}
\end{equation}
with the rotation operator
\begin{eqnarray}
\hat R(\Omega) = e^{-\imath \alpha \hat J_z} e^{-\imath \beta \hat J_y}
e^{-\imath \gamma \hat J_z}~~~.\label{rotop}
\end{eqnarray}
Here, $''\Omega''$ represents a set of Euler angles 
($\alpha, \gamma = [0,2\pi],\, \beta= [0, \pi]$) and the 
$\hat{J}^{,}s$ are angular-momentum operators. 

In the present work we have employed $N=3,4,5$ ($2,3,4$)  for neutrons
(protons). The two valence neutrons (protons)  are occupying N= 4 (3) and 5 (4) shells that give rise to the negative parity states. The deformations used to
generate the Nilsson basis configuration are given in Table \ref{tab1} and have been adopted from earlier works \cite{JS21,Chanli15,moller08}. The Nilsson intrinsic states are then projected onto the states with good angular-momentum through  three-dimensional projection.

\begin{figure}[htb]
 \centerline{\includegraphics[trim=0cm 0cm 0cm
0cm,width=0.55\textwidth,clip]{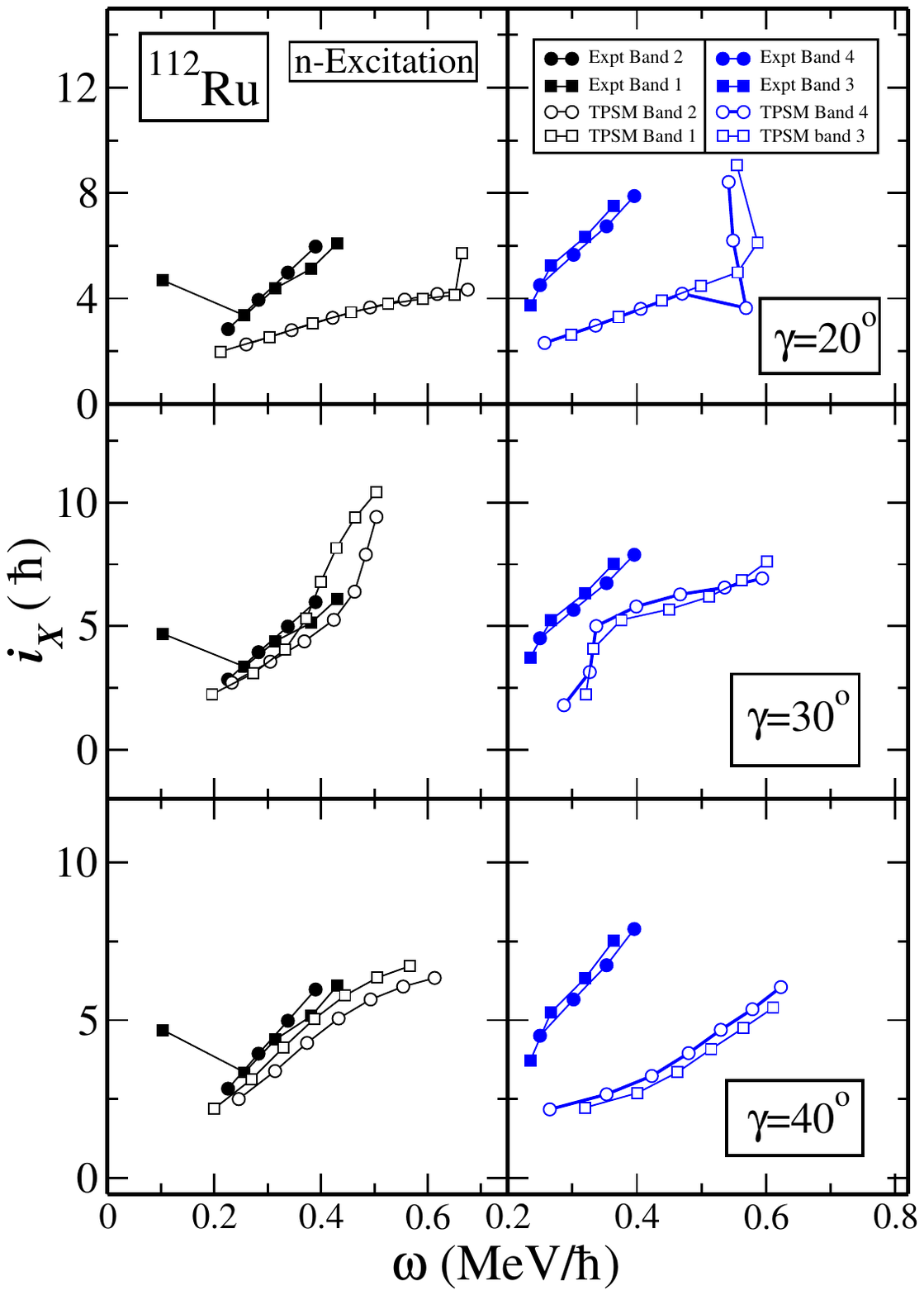}} \caption{(Color
   online) Comparison of the aligned angular momenta,
  $i_X =i_x(\omega)-i_{x,ref}(\omega)$, where $\hbar \omega = \frac{E_{\gamma}}{I^i_{x}(\omega)-I^f_{x}\omega}$,$I_x\omega = \sqrt{I(I+1)-K^2}$ and $i_{x,ref}(\omega)=\omega(J_0 +\omega^2 J_1)$. The reference band Harris parameters
used are $J_0 = 14$ and $J_1 = 15$, obtained from the measured energy
levels as well as those calculated from the TPSM results
for neutron excitation in  $^{112}$Ru.
   }
\label{alignneut112gamma}
\end{figure}
The projected basis states of Eq.~(\ref{basis})
are then used to diagonalize the shell model Hamiltonian. As in our earlier studies, we have employed the pairing plus quadrupole-quadrupole Hamiltonian \cite{JG12,bh14}
\begin{equation}
\hat H = \hat H_0 - {1 \over 2} \chi \sum_\mu \hat Q^\dagger_\mu
\hat Q^{}_\mu - G_M \hat P^\dagger \hat P - G_Q \sum_\mu \hat
P^\dagger_\mu\hat P^{}_\mu .
\label{hamham}
\end{equation}
The corresponding triaxial Nilsson Hamiltonian is the mean-field of
the above model Hamiltonian and is given by:
\begin{equation}
\hat H_N = \hat H_0 - {2 \over 3}\hbar\omega\left\{\epsilon\hat Q_0
+\epsilon'{{\hat Q_{+2}+\hat Q_{-2}}\over\sqrt{2}}\right\}.
\label{nilsson}
\end{equation}
In the above equation, $\hat H_0$ is the spherical single-particle
Nilsson Hamiltonian \cite{Ni69}. The monopole pairing strength $G_M$ 
of the standard form
\begin{figure}[htb]
\centerline{\includegraphics[trim=0cm 0cm 0cm
0cm,width=0.55\textwidth,clip]{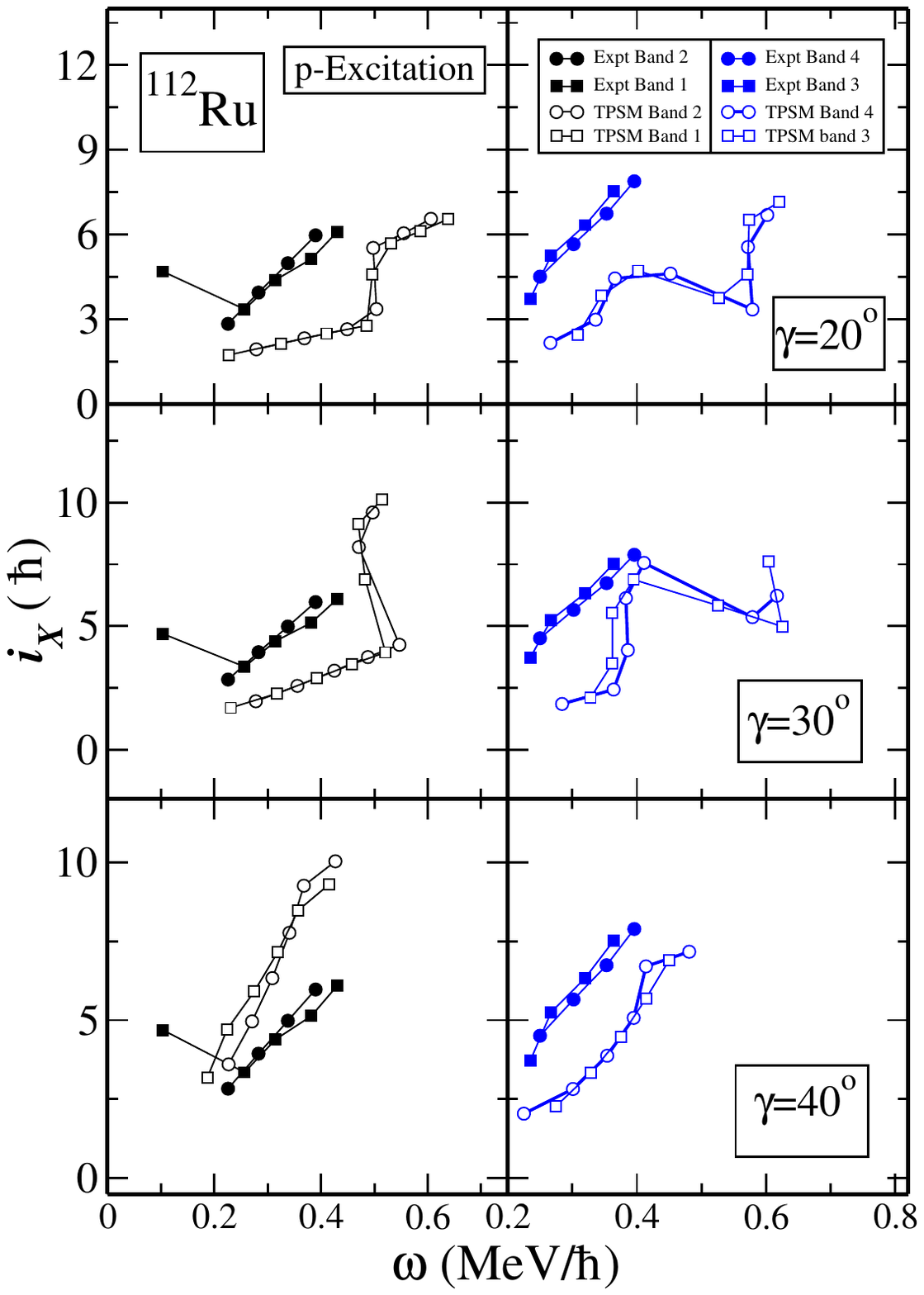}} \caption{(Color
   online) Comparison of the aligned angular momenta,
   $i_X =i_x(\omega)-i_{x,ref}(\omega)$, where $\hbar \omega = \frac{E_{\gamma}}{I^i_{x}(\omega)-I^f_{x}\omega}$,$I_x\omega = \sqrt{I(I+1)-K^2}$ and $i_{x,ref}(\omega)=\omega(J_0 +\omega^2 J_1)$. The reference band Harris parameters
used are $J_0 = 14$ and $J_1 = 15$, obtained from the measured energy
levels as well as those calculated from the TPSM results
for proton excitation in  $^{112}$Ru.
  }
\label{alignprot112gamma}
\end{figure}
\begin{eqnarray}
G_M = {{(G_1 \mp G_2{{N-Z}\over A})}\frac{1}{ A}} (MeV).\label{pairing}
\end{eqnarray}
\begin{figure}[htb]
 \centerline{\includegraphics[trim=0cm 0cm 0cm
0cm,width=9.5cm,height=8cm,clip]{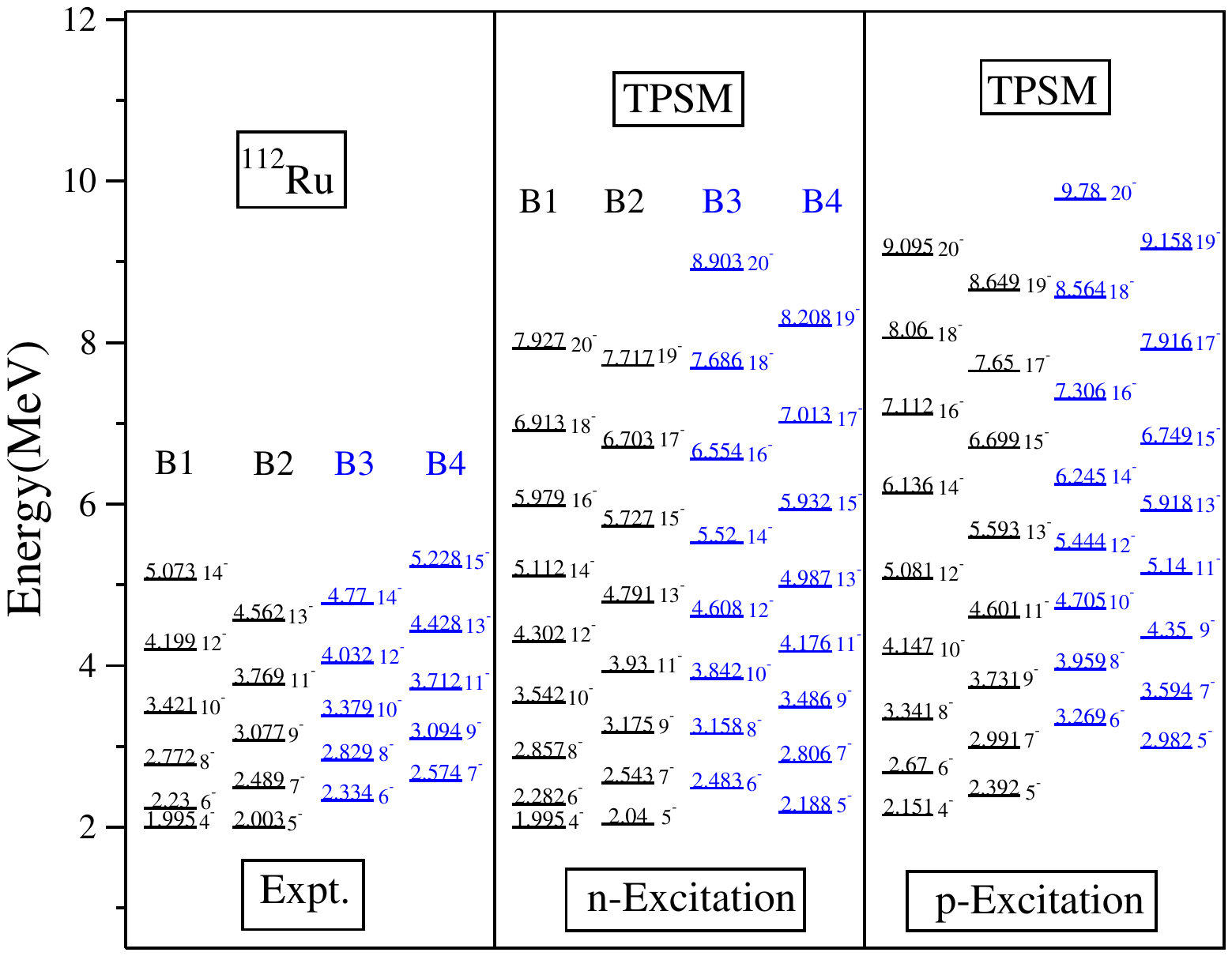}} \caption{(Color
   online) TPSM projected energies after configuration mixing for both neutron- and proton-excitations are compared with experimental data \cite{NPRU} for $^{112}$Ru isotope.
  }
\label{fig:112E}
\end{figure}
In the present calculation, we considered $G_1=22.68$ and $G_2=16.22$, which approximately reproduce the observed odd-even mass difference in the studied mass region.
The quadrupole pairing strength $G_Q$ is
assumed to be proportional to $G_M$, and the proportionality
constant being fixed as 0.18. These interaction strengths are
consistent with those used earlier for the same mass region
\cite{Chanli15,Chanli16}.

The projected TPSM  wavefunction is then given by
\begin{equation}
|\sigma, I M \rangle = \sum_{K \kappa} f^\sigma_{K\kappa} \hat{P}^I_{MK} | \phi_\kappa \rangle .
\label{wave1} 
\end{equation}
Here, the index $\sigma$ labels the states with same angular momentum and $\kappa$
the basis states. In Eq.~(\ref{wave1}), $f^{\sigma}_{K\kappa} $ are the expansion coefficients of the wavefunction in terms
of the non-orthonormal projected basis states, $ \hat{P}^I_{MK} | \phi_\kappa \rangle $. These coefficients are not
probability amplitudes in the usual quantum mechanical sense and in the work of
ref.~\cite{WANG2020}, modified expansion coefficients $(g^{\sigma}_{K\kappa})$ are defined which
are expanded in terms of an orthonormal basis set in the following manner :
\begin{equation}
g^{\sigma}_{K\kappa} = \sum_{K^\prime\kappa^\prime} f^\sigma_{K^\prime\kappa^\prime}\langle K\kappa| \hat{P}^I_{MK^\prime} | \phi_{\kappa^\prime} \rangle =\sum_{K^\prime\kappa^\prime} f^\sigma_{K^\prime\kappa^\prime}N^{1/2 }_{K\kappa K^\prime \kappa^\prime}
\label{orthonormal} 
\end{equation}
where $N$ is the norm matrix and $|K\kappa \rangle $ is an orthonormal basis set.

Finally, the minimization of the projected energy with respect to the expansion coefficient,
$f^\sigma_{K\kappa}$, leads to the Hill-Wheeler type equation
\begin{equation}
\sum_{\kappa '} (H_{\kappa \kappa'} - E_\sigma N_{\kappa \kappa'} ) f^{\sigma}_{
K\kappa'} = 0 ,
\end{equation}
where the normalization is chosen such that
\begin{equation}
\sum_{\kappa \kappa'}f^\sigma_{K\kappa} N_{\kappa \kappa'} f^{\sigma'}_{K\kappa'}
= \delta_{\sigma \sigma'}.
\end{equation}
The above equations are then solved to obtain the energies and the wavefunctions \cite{KY95}.

In the present work, we have also evaluated the transition
probabilities using the TPSM wavefunctions with  the effective charges of $0.5e$ and $1.5e$ for neutrons and
protons. The details of the
transition probability calculations with explicit expressions are given in the review
article \cite{SGH16}.


\section{Results and Discussion}

In comparison to the positive parity bands, there have been only a few theoretical studies to investigate
the negative parity bands in Ru-isotopes and other isotopes in the A $\sim$ 110 mass region. The band heads of the
two-quasiparticle structures have been studied using the D1S Gogny force \cite{NPRU2}, and it has been discussed
that two-proton and two-neutron bands for Ru-isotopes are at a similar excitation energy of about 2 MeV. In the self-consistent
constrained cranking Skyrme calculations with particle number conserving pairing \cite{Dai19}, it
has been shown that calculated moments of inertia
of two-neutron quasiparticle configuration are in better agreement with the experimental data as compared to the two-proton configuration
for $^{108,110,112}$Ru isotopes, and the observed bands have been characterized as neutron excited bands.
\begin{figure}[htb]
 \centerline{\includegraphics[trim=0cm 0cm 0cm
0cm,width=0.55\textwidth,clip]{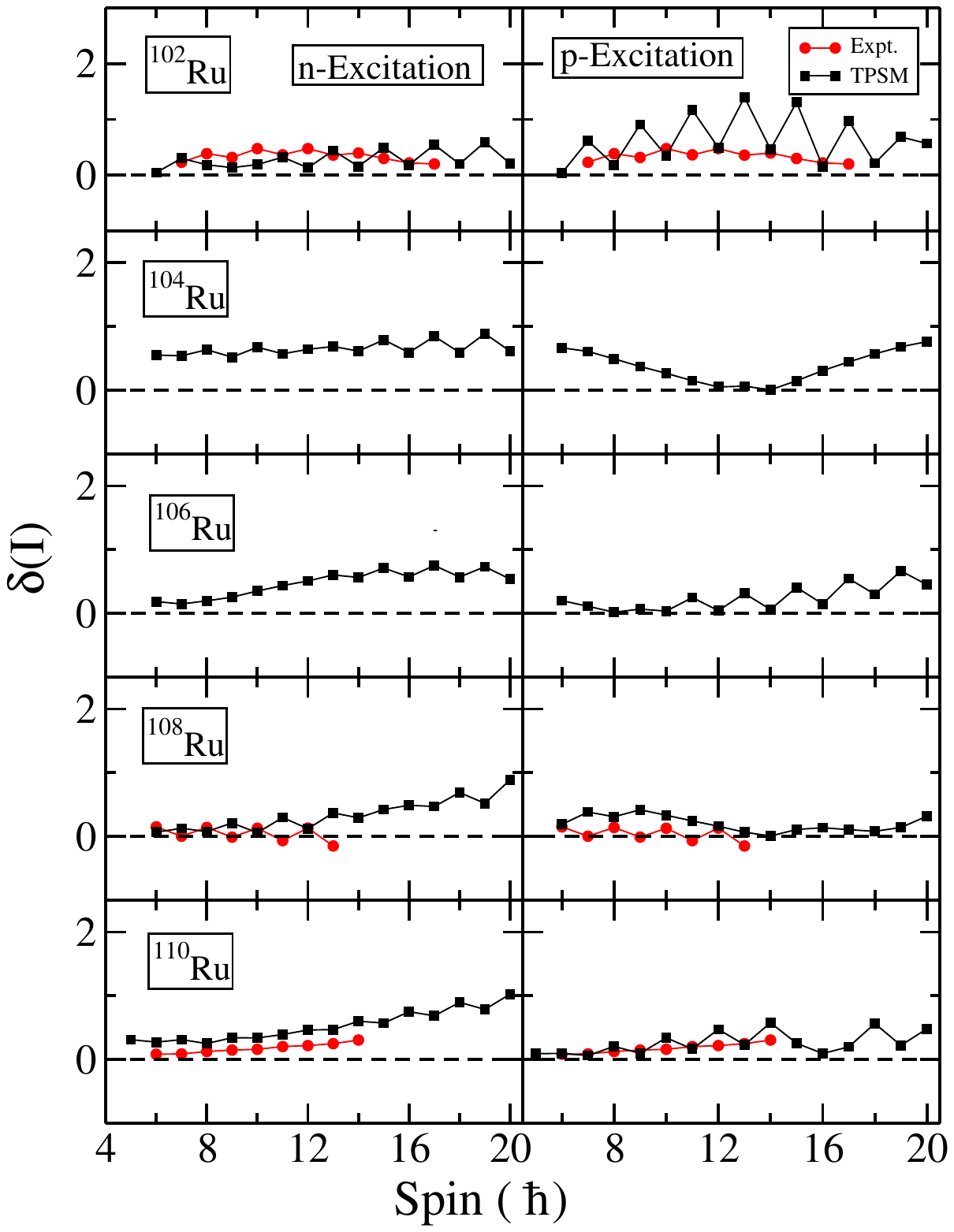}}\caption{(Color online) Energy difference between negative-parity yrast- and yrare- bands for same spin, $I$, $\delta(I)=(E_2(I)-E_1(I))$ in $^{102,104,106,108,110}$Ru.
 }
\label{fig:del}
\end{figure}

\begin{figure}[htb]
 \centerline{\includegraphics[trim=0cm 0cm 0cm
0cm,width=9cm,height=16cm,clip]{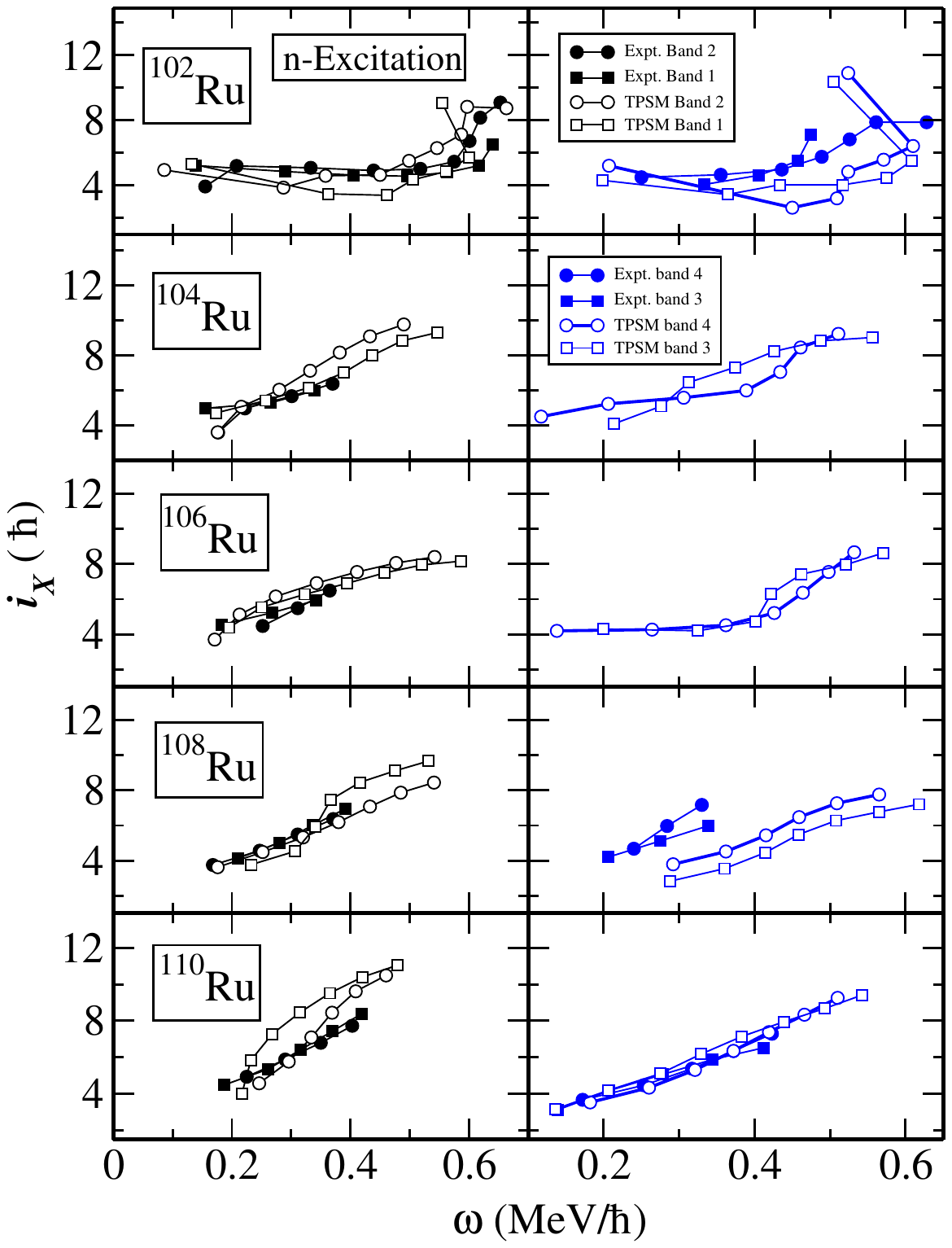}} \caption{(Color
   online) Comparison of the aligned angular momenta,
   $i_X =i_x(\omega)-i_{x,ref}(\omega)$, where $\hbar \omega = \frac{E_{\gamma}}{I^i_{x}(\omega)-I^f_{x}\omega}$,$I_x\omega = \sqrt{I(I+1)-K^2}$ and $i_{x,ref}(\omega)=\omega(J_0 +\omega^2 J_1)$. The reference band Harris parameters
used are $J_0 = 14$ and $J_1 = 15$, obtained from the measured energy
levels as well as those calculated from the TPSM results
for neutron excitation in $^{102,104,106,108,110}$Ru.
   }
\label{fig:k}
\end{figure}
\begin{figure}[htb]
 \centerline{\includegraphics[trim=0cm 0cm 0cm
0cm,width=9cm,height=16cm,clip]{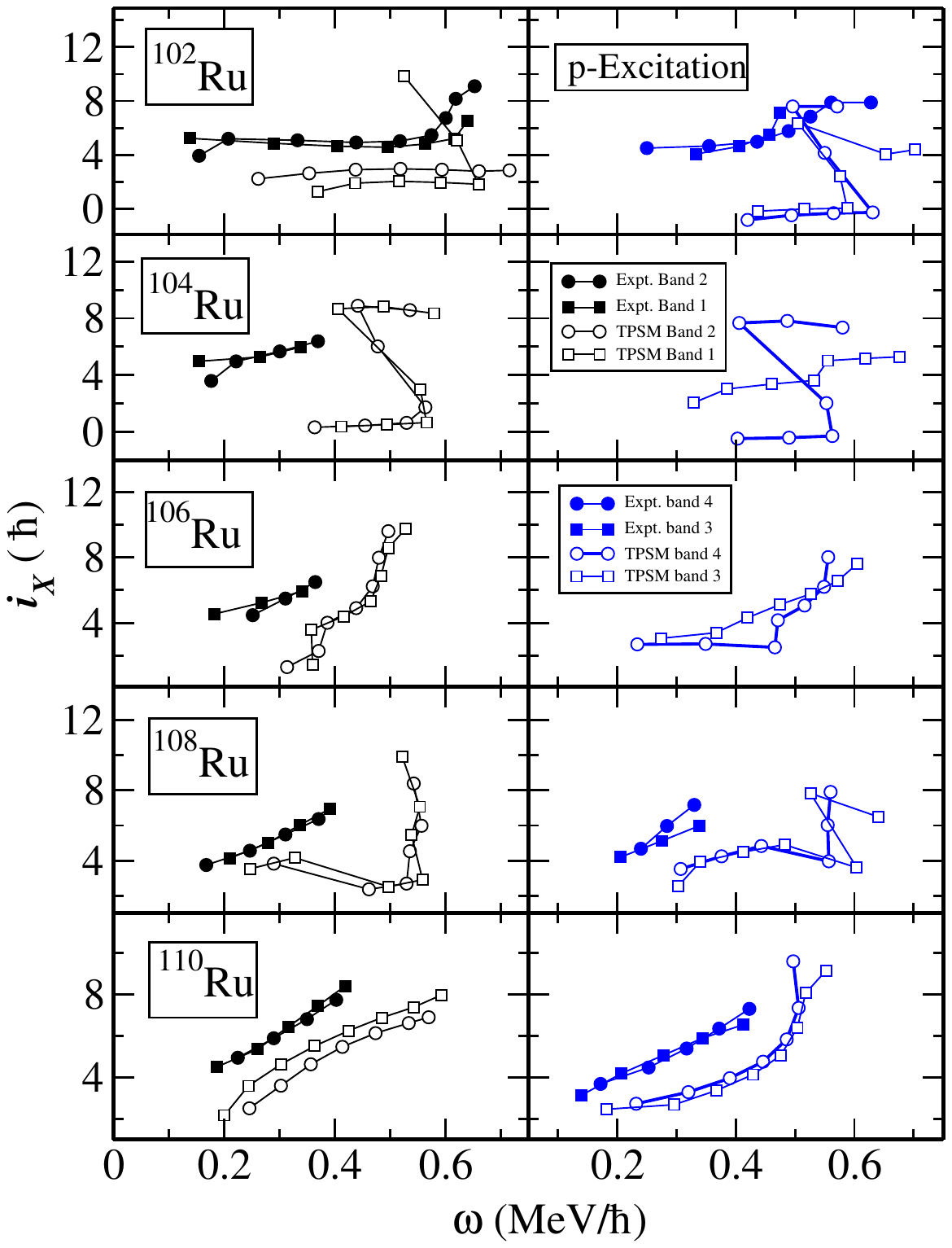}} \caption{(Color
   online) Comparison of the aligned angular momenta,
   $i_X =i_x(\omega)-i_{x,ref}(\omega)$, where $\hbar \omega = \frac{E_{\gamma}}{I^i_{x}(\omega)-I^f_{x}\omega}$,$I_x\omega = \sqrt{I(I+1)-K^2}$ and $i_{x,ref}(\omega)=\omega(J_0 +\omega^2 J_1)$. The reference band Harris parameters
used are $J_0 = 14$ and $J_1 = 15$, obtained from the measured energy
levels as well as those calculated from the TPSM results
 for proton excitation in $^{102,104,106,108,110}$Ru.
   }
\label{fig:ak}
\end{figure}

In order to perform the TPSM study of the negative parity bands, the input parameters
required are deformation values and the strengths of the monopole and the
quadrupole pairing interaction terms. It is expected that deformation of the negative parity
two-quasiparticle states will be slightly different from the yrast positive parity band structures. However,
in the absence of any systematic study of the deformation properties of these band structures, we have adopted the axial deformation values of the ground-state bands from theoretical studies using microscopic-macroscopic model
predictions \cite{moller08} with slight adjustments as in our previous studies \cite{SGH16,Musangu18}. The non-axial deformation values have been varied to reproduce the observed properties of these bands. The deformation values adopted
in the present analysis are listed in Table \ref{tab1}.

The pairing parameters are clearly expected to be different from the
ground-state values as these are two-quasiparticle states, and it is known that pairing is reduced for the excited quasiparticle
states. However, it is difficult to study the reduction in the pairing correlations for the quasiparticle states as in the
BCS approximation, the pairing collapses for the blocked  states and it is imperative to perform the particle-number
projected analysis before variation \cite{sheikh02}.  In the present work, we have investigated the sensitivity of the results on the pairing correlations by varying the
monopole pairing strengths.

We have considered $^{112}$Ru as an illustrative example to investigate the deformation and
pairing dependence of the TPSM results. The
pairing strength parameters that best reproduce the experimental data of
$^{112}$Ru are then used to perform the TPSM calculations for other
Ru-isotopes from A=102 to 110. The reason that this system has been chosen is because doublet band structures
have been observed for this
nucleus up to quite high-spin and the system is well deformed \cite{NPRU}. For other isotopes, for instance,
$^{102}$Ru the data is also available up to quite high-spin. However, this system has vibrational character in the low-spin
region \cite{NPRU3} and the application of TPSM approach becomes unreliable as a single deformed mean-field solution
is adopted in this model.
\begin{figure}[htb]
 \centerline{\includegraphics[trim=0cm 0cm 0cm
0cm,width=9.3cm,height=16cm,clip]{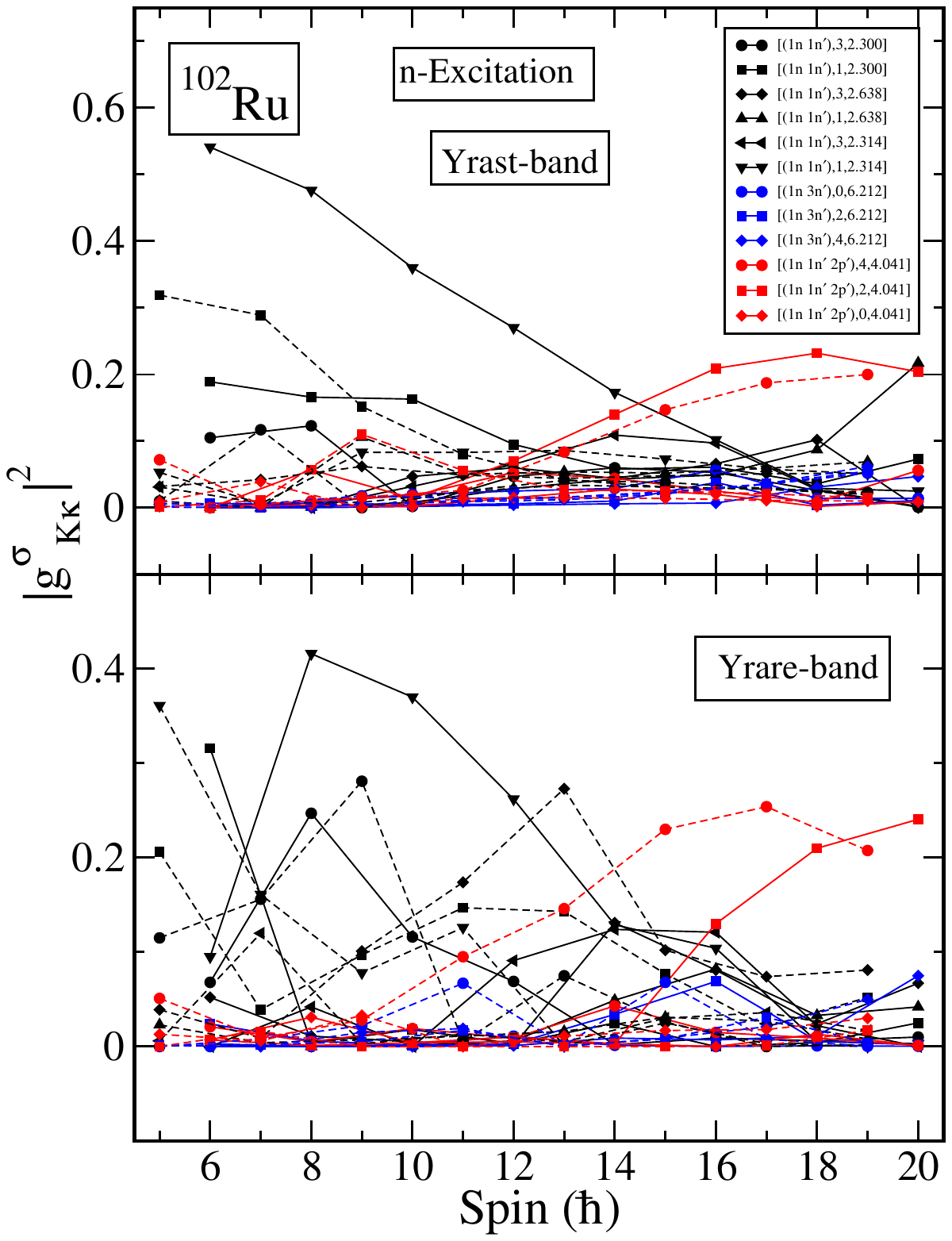}} \caption{(Color
   online) Probabilities of various projected K configurations in the orthonormal wavefunctions of the negative-parity yrast- and yrare- bands after diagonalization for $^{102}$Ru. The curves are labeled by three quantities: quasiparticle character, $''$K$''$ quantum number and energy of the quasiparticle state. For instance, [($1\text{n}1\text{n}^{\prime})$,$3$,$2.3$] designates two--quasineutron state with  K$=3$  having intrinsic energy of $2.3$ MeV. Even- and odd-spin states correspond to $\alpha=0$ (solid lines) and  $\alpha=1$ (dashed lines), respectively.
  }
\label{wavefunction}
\end{figure}

\begin{figure}[htb]
 \centerline{\includegraphics[trim=0cm 0cm 0cm
0cm,width=9.5cm,height=8cm,clip]{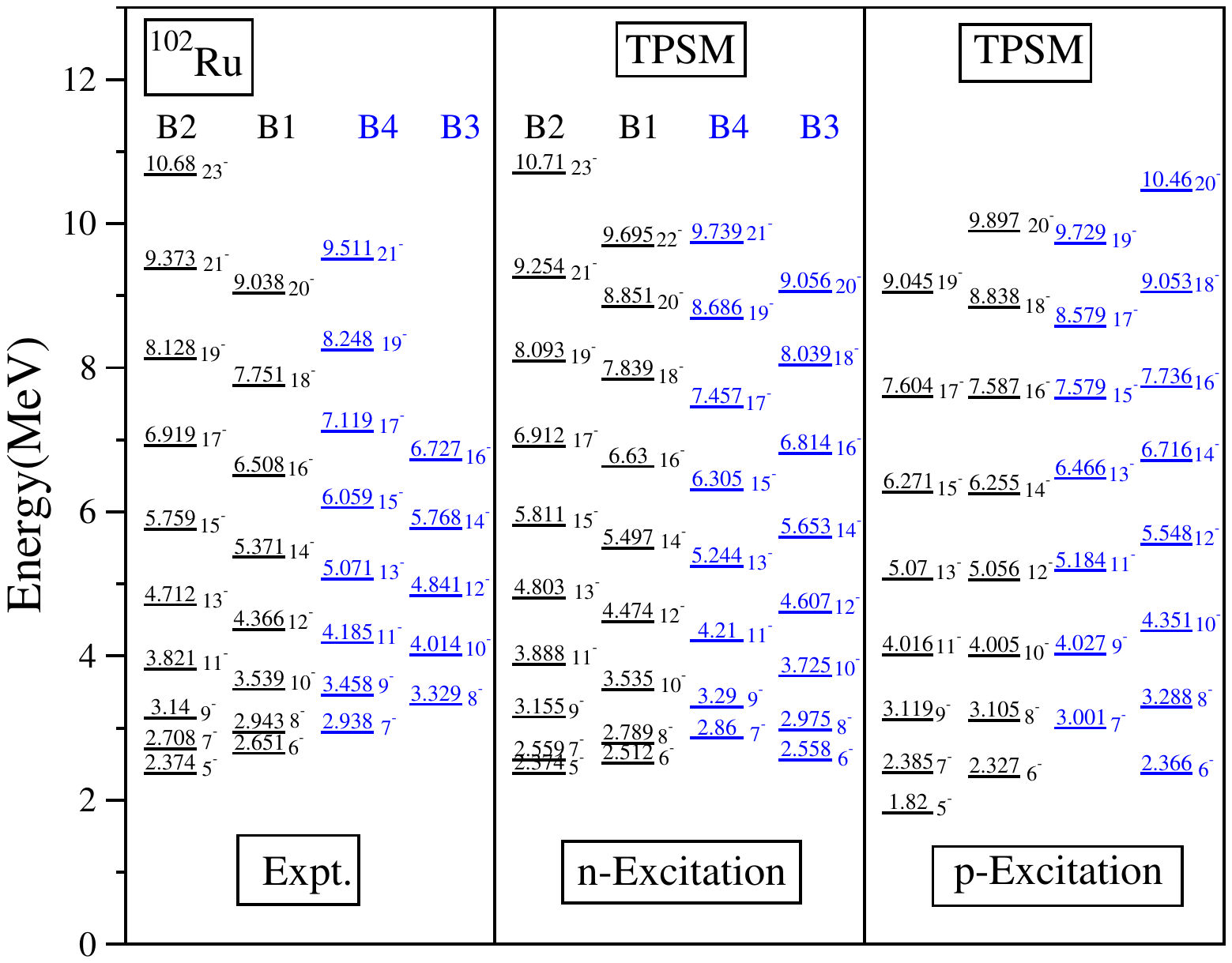}} \caption{(Color
   online) TPSM projected energies after configuration mixing for both neutron- and proton-excitations  are compared with experimental data \cite{NPRU3} for $^{102}$Ru isotope.
  }
\label{fig:102E}
\end{figure}

It has been proposed that the doublet band structures observed in $^{112}$Ru originate from the chiral symmetry breaking mechanism \cite{NPRU}
as the difference of the energies between the two bands, $\delta (I) = (E_2(I) -E_1(I))$, is very small. In Fig.~\ref{del112}, this difference is plotted for different values of pair-gaps ($\Delta$) and non-axial deformation. The differences in the excitation energies of both  two-neutron (left panel) and
two-proton (right panel) quasiparticle configurations are plotted. The results are depicted
only for three representative values of the pair-gaps, $\Delta_n=0.6~\textrm{MeV},~ \Delta_p=1.2~\textrm{MeV}$;~ $\Delta_n=0.6~\textrm{MeV},~ \Delta_p=0.6~\textrm{MeV}$ and $\Delta_n=1.2~\textrm{MeV},~\Delta_p=0.6~\textrm{MeV}$. [ The TPSM calculations
have also been performed for other values of the pair-gaps between 0.6 and 1.2, and the results are not very different from the
three cases depicted in Fig.~\ref{del112}]. It is evident from the results that $\gamma = 30^0$ leads to lowest differences in the energies and agrees
with the corresponding experimental numbers. Further, the results with the pairing set of $\Delta_n=\Delta_p=0.6$~MeV appears to be in
better agreement with the data for the neutron excitation case as compared to the proton excitation.

To further examine the optimum pairing set and the non-axial deformation parameter that reproduce the experimental data
more accurately, we have evaluated the aligned angular
momentum values, $i_x$, for the doublet bands and the results are presented in Figs.~\ref{alignneut112}, \ref{alignprot112}, \ref{alignneut112gamma} and \ref{alignprot112gamma}. As compared to the energies, $i_x$
is  sensitive to the single-particle states occupied by the excited particles and should provide a better
estimate of the optimum pairing and deformation set. The calculated $i_x$ for the neutron excited configuration are shown in Fig.~\ref{alignneut112} for three different
pairing sets, but with the same non-axial deformation parameter of $\gamma=30^\circ$.
It is evident from the figure that the pairing set of $\Delta_n=\Delta_p=0.6~\textrm{MeV}$ provides a better representation of the experimental values as compared to the other two sets, in  particular, $i_x$ for Bands 1 and 2 is reproduced
remarkably well with this set. Further, the slope of the alignment curve, which is the moment of inertia,
is also in good agreement with the data for this set. The alignment calculated with the proton
excitation is depicted in Fig.~\ref{alignprot112} and it is noted that
none of the parameter set is able to reproduce the experimental values. The calculated $i_x$ values depict backbending phenomena, whereas the
experimental values show a smooth increase with spin for both the bands. In Figs.~\ref{alignneut112gamma} and \ref{alignprot112gamma}, the alignments are displayed
for different values of the non-axial deformation parameters and with the pairing set of
$\Delta_n=0.6~\textrm{MeV},~ \Delta_p=0.6~\textrm{MeV}$. It is evident from these figures that $\gamma=30^\circ$ for the
neutron excitation shows the best agreement with the data.


The lowest two negative parity bands obtained for $^{112}$Ru after diagonalization of the shell model Hamiltonian,
for both neutron and proton excitations, are compared with the experimental energies in Fig.~\ref{fig:112E}. It has been
already stated that TPSM Hamiltonian, in the absence of the exchange terms, does not mix neutron and proton excitations and the
two basis spaces can be diagonalized separately. The calculated band structures obtained with the neutron excitation are 
lower in energy as compared to the band structures with the proton excitation. The lowest negative band structure (labelled
as B1 and B2 in Fig.~\ref{fig:112E}) is quite reasonably reproduced by the TPSM  calculations with neutron excitation, and the deviation for the
highest spin, I=14, for this band is 0.039 MeV. For the excited band (labelled as B3 and B4), significant deviations are noted for most of the spin states and for the highest spin observed, I=15, the calculated value has a deviation of 0.704 MeV.

The alignment plotted in Fig.~\ref{alignneut112}
also depicts significant deviations, although the slope is in agreement. The origin of this deviation is not evident at this stage and
further analysis is needed, for instance, using a self-consistent mean-field approach. 
In Fig.~\ref{fig:112E}, we have also
provided the energies of the proton excited bands as in future experimental studies more band structures will be populated and some of them may
correspond to the proton excitation.

It needs to be mentioned that TPSM energies in Fig.~\ref{fig:112E}
have been plotted for lower-spin values as well, which are not known experimentally. The reason is that the studied
negative-parity band structures have dominant K=2 and 3 configurations as is evident from the TPSM wavefunctions
and, therefore, the band structures will have either I=2 or 3 as bandheads. The yrast negative-parity band has been plotted
from the spin-value observed in the experimental data as calculated TPSM energy for this state is set equal to the corresponding
experimental value. For the negative-parity yrare-band, we have also provided a few low-lying states which are not known experimentally.
The reason that these low-lying states are not known is probably because these are mixed with spherical states as many negative
parity states are observed in these nuclei which are not members of the rotational bands.  

\begin{figure}[htb]
 \centerline{\includegraphics[trim=0cm 0cm 0cm
0cm,width=9.5cm,height=8cm,clip]{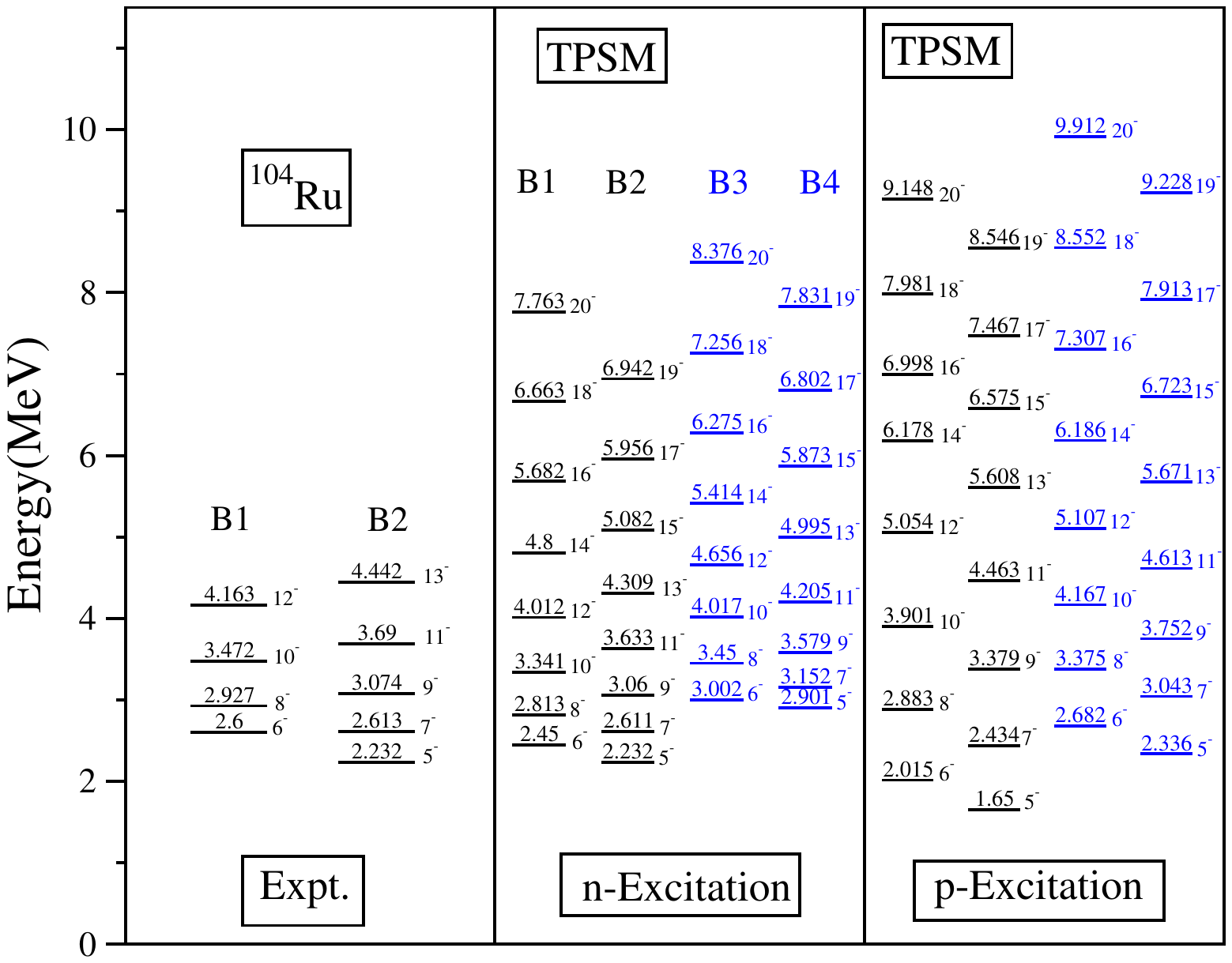}} \caption{(Color
   online) TPSM projected energies after configuration mixing  for both neutron- and proton-excitations are compared with experimental data \cite{NPRU2} for $^{104}$Ru isotope.
  }
\label{fig:104E}
\end{figure}

\begin{figure}[htb]
 \centerline{\includegraphics[trim=0cm 0cm 0cm
0cm,width=9.5cm,height=8cm,clip]{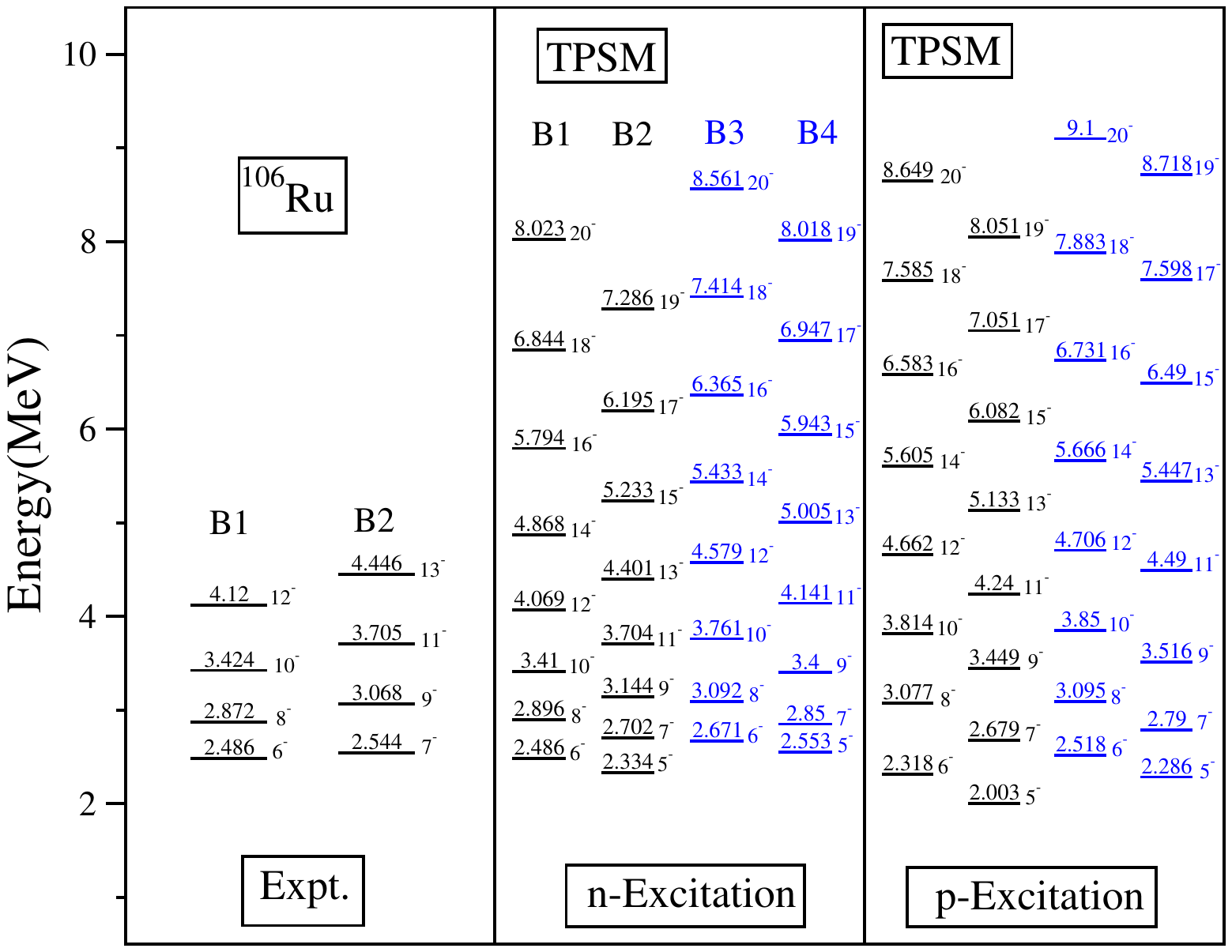}} \caption{(Color
   online) TPSM projected energies after configuration mixing  for both neutron- and proton-excitations are compared with experimental data \cite{NPRU2} for $^{106}$Ru isotope.
  }
\label{fig:106E}
\end{figure}

\begin{figure}[htb]
 \centerline{\includegraphics[trim=0cm 0cm 0cm
0cm,width=9.5cm,height=8cm,clip]{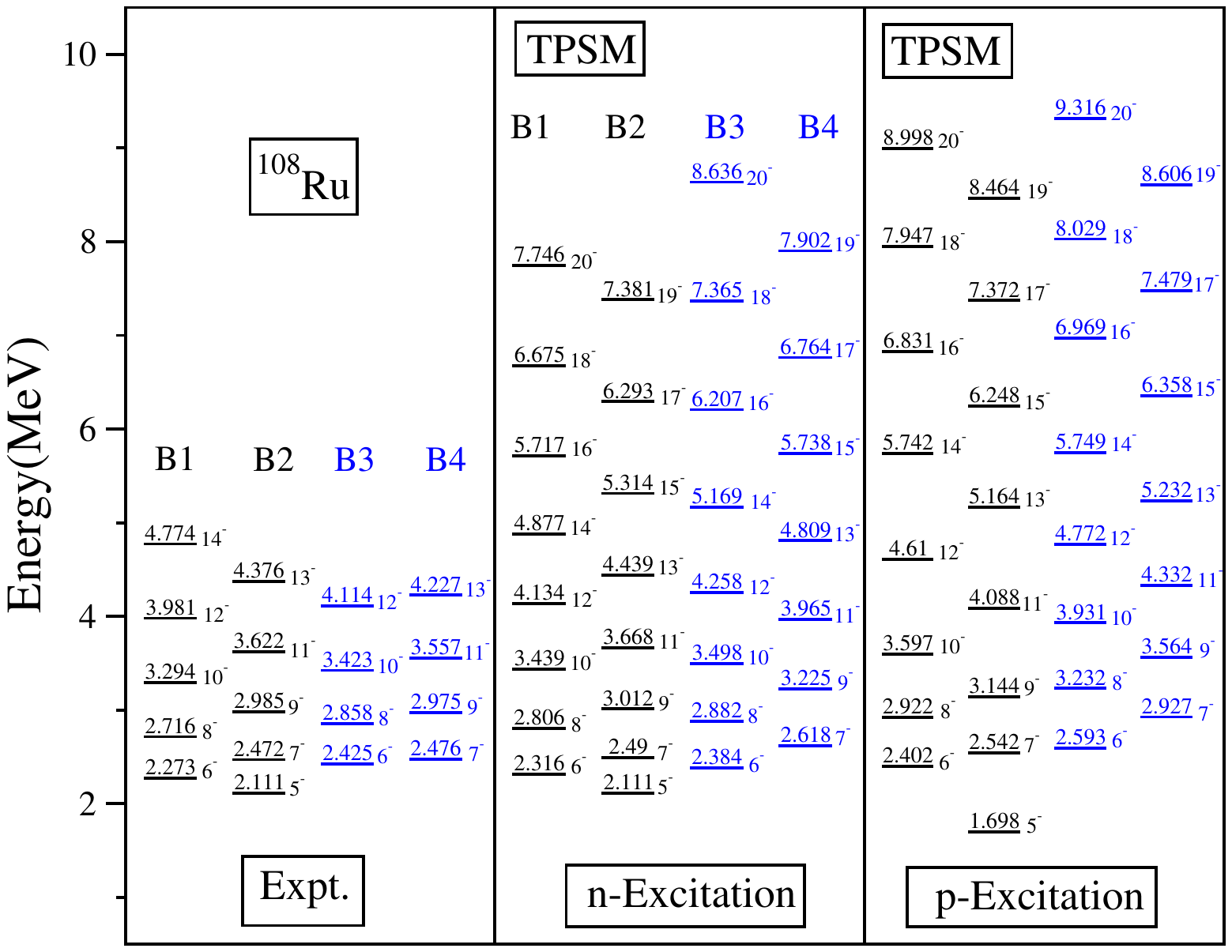}} \caption{(Color
   online) TPSM projected energies after configuration mixing  for both neutron- and proton-excitations are compared with experimental data \cite{NPRU} for $^{108}$Ru isotope.
  }
\label{fig:108E}
\end{figure}
We have performed TPSM study for other Ru-isotopes from A=102 to 110 with the axial and non-axial deformation values listed in Table \ref{tab1}. The pairing strengths are same as those adjusted
to reproduce the properties of $^{112}$Ru. The results of the energy difference between the two doublet band, $\delta (I)$,  are displayed
in Fig.~\ref{fig:del} for both proton and neutron excitations. Neutron excitation energies are slightly lower than the corresponding
proton energies, however, both are compared with the available experimental data as the proton excitation spectra can become
favoured with slight adjustments in the pairing and deformation parameters. It is noticed from the figure that TPSM calculated $\delta(I)$ from neutron excitation energies agrees well with the observed energies of $^{102}$Ru, $^{108}$Ru and $^{110}$Ru.  It is also evident from the figure that $\delta(I)$ for the lowest two proton bands also agrees with the data, except for $^{102}$Ru which shows a staggering pattern. For other isotopes yrare-band has not been observed.

The alignments of the isotopes are plotted in Fig.~\ref{fig:k} for neutron excitation spectra and compared with the corresponding experimental numbers, wherever available. It is quite remarkable to note from the figure that $i_x$ values are reproduced  well for all the isotopes. For $^{102}$Ru, upbend is observed for bands B1 and B2 at $\hbar \omega \approx 0.55$ MeV and is well described by the TPSM calculations. For other isotopes, $i_x$ depicts a smooth increase with rotational frequency and is easily understood as unpaired particles align towards the rotational axis. For $^{106}$Ru, TPSM calculated bands B3 and B4 show upbends, but there is no experimental data to confirm this band crossing phenomenon. The alignments for the proton excitation bands are displayed in Fig.~\ref{fig:ak} and it is noted that in most of the cases, the band crossing is expected as either an upbend or a backbend is observed. These alignments considerably differ with the experimental values.

To examine the nature of the bandcrossing phenomenon observed in $^{102}$Ru, the wavefunction probabilities are displayed
in Fig.~\ref{wavefunction} for both negative-parity yrast- and yrare- bands. It is noted from the figure that for the
negative-parity yrast-band, the dominant component before I=14
is a mixture of many $(1n1n')$ configurations. After I=14, the dominant component in the
wavefunction is a four quasiparticle state, $(1n1n'2p')$ having K=2, which is a two-proton
aligned state built on the basic $(1n1n')$ configuration. It is observed from   Fig.~\ref{wavefunction} that there
is a considerable mixing between the two-quasiparticle and the proton-aligned four-quasiparticle states and,
therefore, upbend rather than a backbend is expected and this is what is seen in the alignment plot for $^{102}$Ru in
Fig.~\ref{fig:k}. For the negative-parity yrare-band, bandcrossing is also expected at I=14 as the wavefunction in the lower
panel of Fig.~\ref{wavefunction} depicts dominant $(1n1n')$ contribution before this spin value and after it, the wavefunction
is dominated by $(1n1n'2p')$ four-quasiparticle state with K=4. 


\begin{figure}[htb]
 \centerline{\includegraphics[trim=0cm 0cm 0cm
0cm,width=9.5cm,height=8cm,clip]{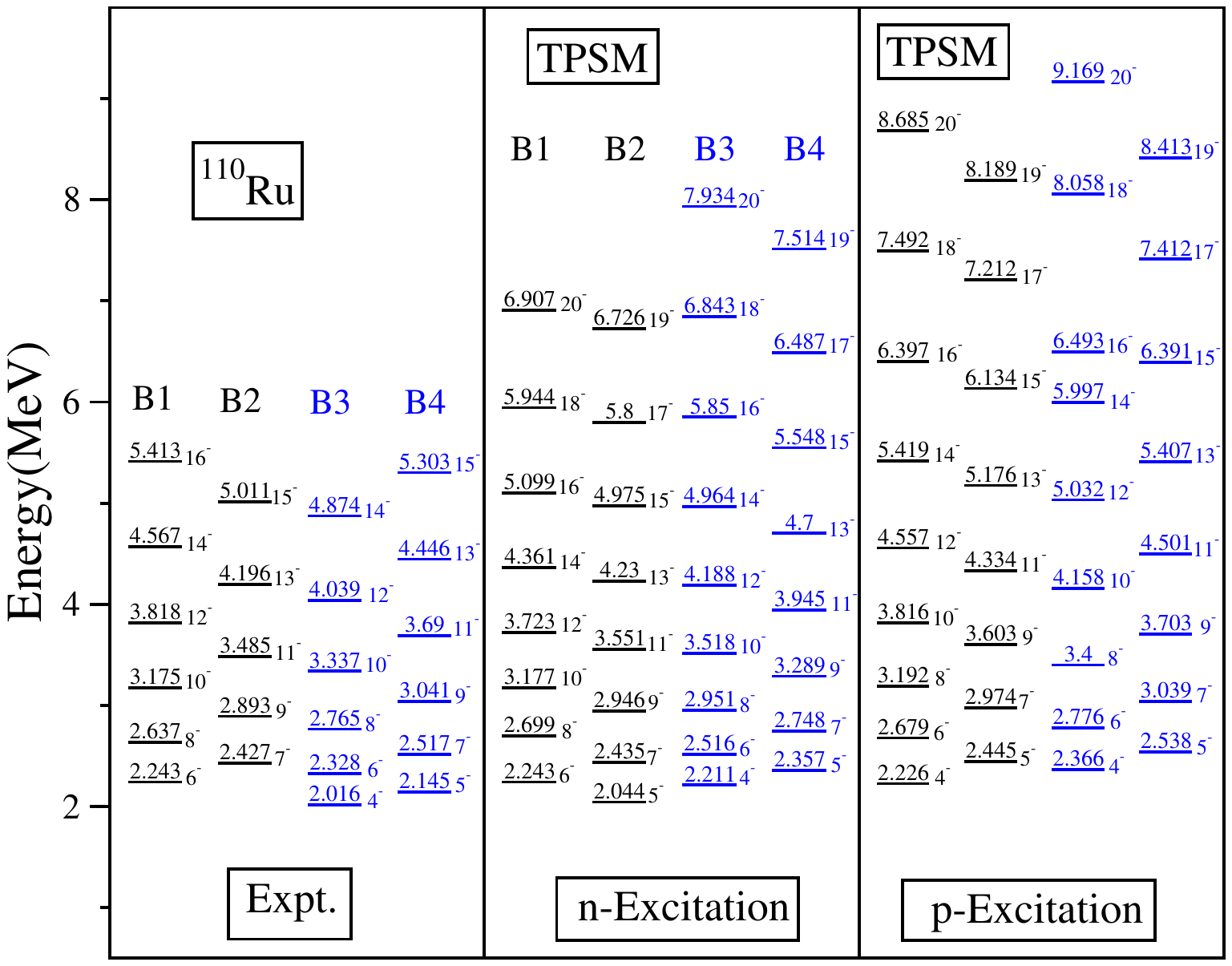}}\caption{(Color
   online) TPSM projected energies after configuration mixing  for both neutron- and proton-excitations are compared with experimental data \cite{NPRU} for $^{110}$Ru isotope.
  }
\label{fig:110E}
\end{figure}

\begin{figure}[htb]
 \centerline{\includegraphics[trim=0cm 0cm 0cm
0cm,width=9.5cm,height=12cm,clip]{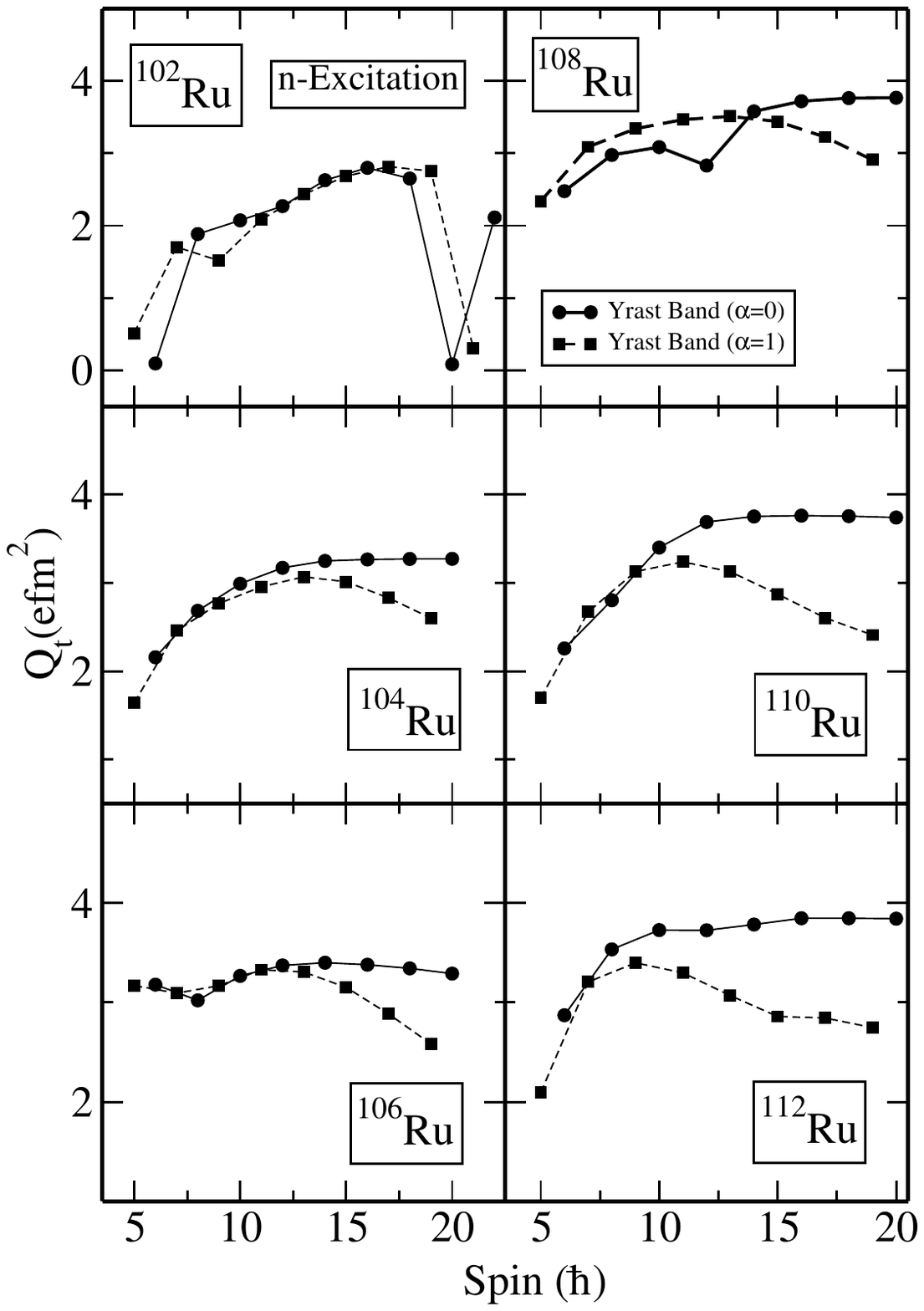}}\caption{(Color
   online) TPSM  calculated transition quadrupole moments, $Q_t (e \textrm{fm}^2)$, for the negative-parity
   yrast-band in $^{102-112}$Ru isotopes.
  }
\label{fig:qtyrast}
\end{figure}

TPSM calculated energy spectra  are compared with the experimental energies in Figs.~\ref{fig:102E}, \ref{fig:104E}, \ref{fig:106E}, \ref{fig:108E} and \ref{fig:110E} for $^{102}$Ru,$^{104}$Ru, $^{106}$Ru, $^{108}$Ru and $^{110}$Ru, respectively. For $^{102}$Ru, the experimental energies are known up to I=23, and it is observed that TPSM calculations with neutron excitation reproduces the data quite well, the deviations for most of the states is less than 0.2 MeV. The proton excitation bands  at low-spin are almost degenerate with the neutron bands, but at higher spins, they become unfavoured. In the case of $^{104}$Ru and $^{106}$Ru, the data is available up to I=13  and only one band is known. The TPSM results are in reasonable agreement with the data for both the nuclei. For $^{108}$Ru
and $^{110}$Ru, doublet negative parity bands have been observed and it is noted that TPSM calculation are able to reproduce the low-lying states in both the nuclei quite well, however, deviations are noted for the excited states.

\begin{figure}[htb]
 \centerline{\includegraphics[trim=0cm 0cm 0cm
0cm,width=9.5cm,height=12cm,clip]{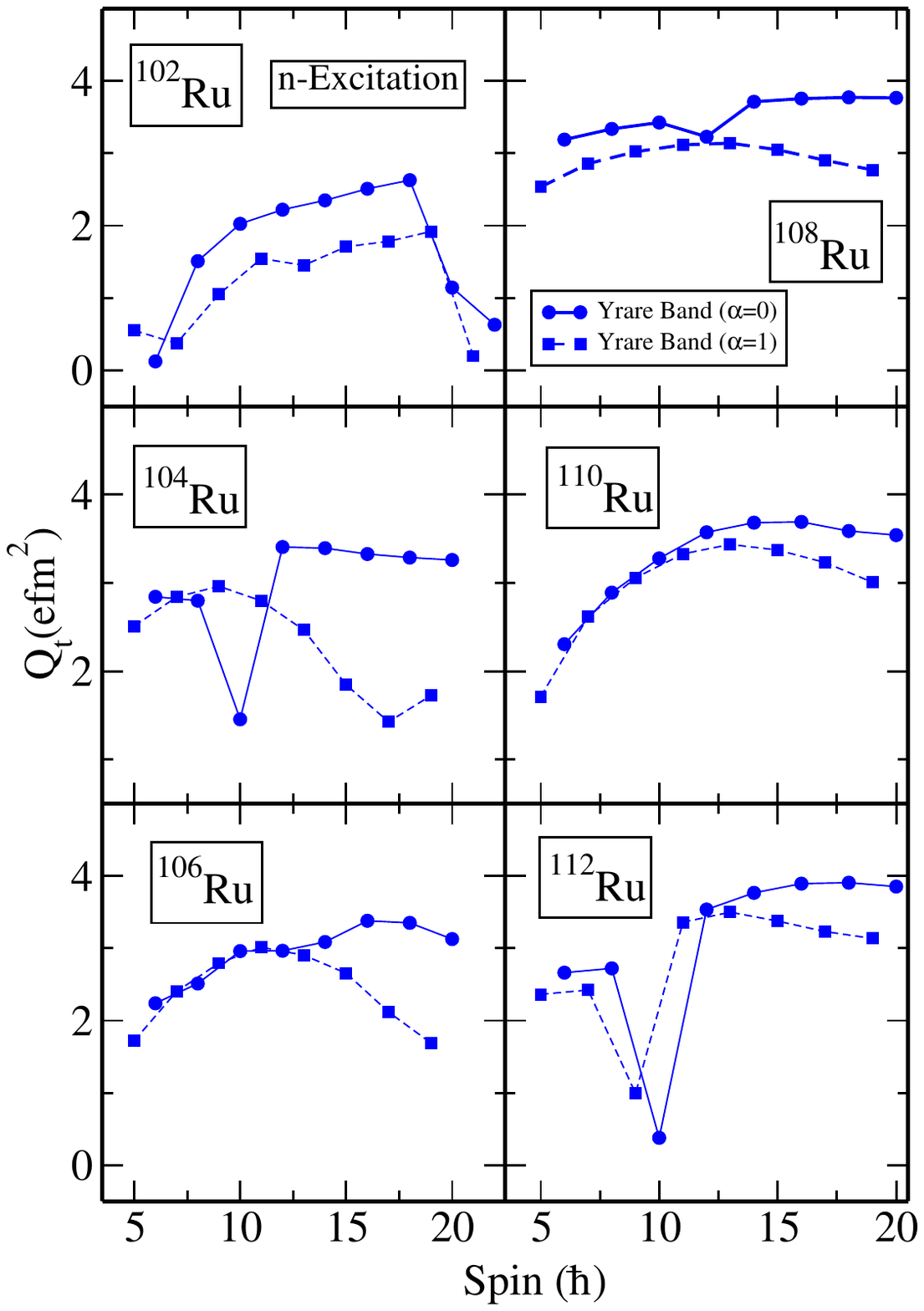}}\caption{(Color
   online)  TPSM calculated transition quadrupole moments, $Q_t (e \textrm{fm}^2)$, for the negative parity
   yrare-band in $^{102-112}$Ru isotopes.
  }
\label{fig:qtyrare}
\end{figure}
As transition probabilities are more sensitive to any structural changes, we have also
evaluated the transition quadrupole moment ($Q_t$) for the negative-parity yrast- and yrare- bands for all the
studied isotopes and the results
are presented in Figs.~\ref{fig:qtyrast} and \ref{fig:qtyrare}. It is evident from the figures that in the bandcrossing region, $Q_t$ depicts
large changes since the wavefunctions are mixed in this region. For $^{102}$Ru, the drop is observed at I=20 and 21 for the two signature
branches and these are the spin values for which large changes are oberved in the aligned angular momentum as seen
from Fig.~\ref{fig:k}. It is noted from Fig.~\ref{fig:qtyrare} that negative parity yrare-band for $^{112}$Ru also depicts a large drop at I=10 and
substantiates the occurence of bandcrossing at this spin value with the aligned angular momenta showing a rapid increase
in Fig.~\ref{alignneut112} for the corresponding angular frequency value.

\section{Summary and Conclusions}

In the present work, the TPSM approach has been extended to investigate
the negative parity bands in even-even systems. In all the previous
versions of the model, the quasiparticle excitations were considered
from a single oscillator shell and it was possible to investigate only
positive parity bands. Here, we have extended the  model space by considering quasiparticle basis from two different
oscillator shells and a detailed investigation has been performed for the
negative parity bands observed in $^{102-112}$Ru as considerable data is available for these isotopes.

Both proton and neutron quasiparticle excitations have been considered in the present analysis and it has been observed that neutron spectra
is slightly lower than the proton one. However, the comparison of alignments clearly delineated the two spectra with neutron alignment observed to be in better agreement with the data and proton alignment shows significant deviations. Further, transition quadrupole moments for both
negative-parity yrast- and yrare- bands have been evaluated and it is noted that these depict large changes
in the bandcrossing region.

In order to improve the predictive power of the present investigation, deformation and pairing parameters need to be
determined for the negative parity bands.
It is quite evident that parameters known for the yrast positive parity configurations cannot be used for the two-quasiparticle negative parity states. It is essential to deduce these parameters from the microscopic models, for instance, the energy
density functional approaches with  the blocking technique \cite{schunch10}. We are planning to perform this study using the Skyrme density functional approach \cite{HFODD21} and these parameters will then be used to evaluate the properties of negative parity band structures
more accurately and will allow us to investigate the chiral symmetry origin proposed for some of the studied isotopes.

In some of the nuclei studied, the octupole deformation is predicted to be an important degree of freedom \cite{dejb95,NPRU3} and TPSM approach needs to be augmented to include the octupole deformation. We are considering to  include the octupole correlations in the TPSM framework. This can be achieved in two phases.
In the first phase, the octupole-octupole interaction will be included in the Hamiltonian with the mean-field having well defined parity. In this way, the octupole correlations will be added as a perturbation correction. In the second phase, the octupole mean-field will be
considered in the Nilsson state with the explicit breaking of the reflection symmetry. This broken symmetry can then be restored using the standard parity projection formalism \cite{EGIDO92,EGIDO91,GARROTE97,garrote98}.

Further, a major deficiency of the present approach is that deformation and pairing fields are kept fixed irrespective of the
angular-momentum and quasiparticle configuration as the projection is carried out after variation.
This is a gross simplification since it is known that mean-field is modified for
multi-quasiparticle and higher angular-momentum states. It is highly desirable that generator coordinate method (GCM)
\cite{hill53,gfiffin57} be developed with the TPSM wavefunctions as basis configurations. In recent years, considerable
progress has been made in the implementation of the GCM related techniques in nuclear physics and microscopic
structure of various phenomena have been explored \cite{VALOR2000,RODRIG02,bender08,yao10,egido16,robledo22}.

\section*{Acknowledgments}
The authors are grateful to Prof. Stefan Frauendorf for illuminating discussions. 
The authors are also thankful to Science and
Engineering Research Board (SERB), Department of Science and
Technology (Govt. of India) for providing financial assistance under the 
Project No.CRG/2019/004960, and for the INSPIRE fellowship to one of the author (NN).

\onecolumngrid
 \appendix
 \renewcommand{\theequation}{A.\arabic{equation}}
 \setcounter{equation}{0}
\section*{Appendix}
The Hamiltonian in terms of proton and neutron degrees of freedom employed in the TPSM approach is given by
\begin{align}
\hat H &= \hat H_0 - {\chi_{pp} \over 2} \sum_\mu \hat Q^\dagger_\mu (p)\hat Q^{}_\mu(p)-{ \chi_{nn} \over 2 }\sum_\mu\hat Q^\dagger_\mu(n)\hat Q^{}_\mu(n)-\chi_{np}\sum_\mu\left( \hat Q^\dagger_\mu(p)
  \hat Q^{}_\mu(n)+\hat Q^\dagger_\mu(n)\hat Q^{}_\mu(p)\right)\nonumber\\
&- G_{M}^{p} \hat P^\dagger_{0}(p) \hat P^{}_{0}(p)
 - G_{M}^{n} \hat P^\dagger_{0}(n) \hat P^{}_{0}(n) -G_{Q}^p \sum_\mu  \hat P^\dagger_\mu(p)\hat P^{}_\mu(p)
 -G_{Q}^n \sum_\mu \hat P^\dagger_\mu(n)\hat P^{}_\mu(n)~,
\label{hamal}
\end{align}
where the labels ``n (p)''  denote neutron (proton) states. The explicit form of the one-body operators in the above equation are given by\\
\begin{eqnarray}
  \hat Q^\dagger_\mu =\sum_{\alpha \beta}Q_{\mu \alpha \beta} c^{\dagger}_\alpha c_\beta~,\hspace{0.5cm} \hat P^\dagger_0 = \frac{1}{2}\sum_{\alpha} c^{\dagger}_\alpha c^{\dagger}_ {\bar{\alpha}}~,\hspace{0.5cm} \hat P^\dagger_\mu
  = \frac{1}{2}\sum_{\alpha \beta} Q_{\mu \alpha \beta} c^{\dagger}_\alpha c^{\dagger}_{\bar{\beta}}~~.
  \end{eqnarray}
Here the quadrupole matrix elements $Q_{\mu \alpha \alpha^\prime} = \delta_{NN^\prime}\langle Njm|Q_\mu|N^\prime j^\prime m^\prime\rangle$ with  $\alpha =\{Njm\}$, $\bar{\alpha}$ represents the time-reversed state of $\alpha$ and the dimensionless mass quadrupole operator is \cite{KY95}
\begin{equation}
 Q_\mu=\sqrt{4\pi\over 5}~{m \omega r^2\over\hbar}Y_{2\mu}~.
\end{equation}  
In the evaluation of the matrix elements of the Hamiltonian of Eq.~(\ref{hamal}), the exchange terms are disregarded. Using
the Wick's theorem to one of the terms in Eq.~(\ref{hamal}) with $|\Phi\rangle$ as the reference state, we have  
\begin{align}
 \hat{O}^\dagger \hat{O}
  ={\langle\Phi| \hat{O}|\Phi\rangle}^{2} + \langle\Phi|  \hat{O}|\Phi\rangle \bigl(:\hat{O}^{\dagger}: + :\hat{O}: \bigl) + :\hat{O}^{\dagger}::\hat{O}:
  =\hat{H}^{(0)}+\hat{H}^{(1)}+\hat{H}^{(2)} \label{HOH1}~~.
\end{align}
Now using the generalized Wick's theorem \cite{balian,HS79,HS80}, we evaluate the matrix elements between the projected quasiparticle states of
Eq.~(\ref{HOH1}). First of all, for the vacuum state, we have
\begin{align}
  \langle\Phi| \hat{H}^{(0)}[\Omega]|\Phi\rangle& ={\langle\Phi| \hat{O}|\Phi\rangle}^2~, \nonumber\\
  \langle\Phi| \hat{H}^{(1)}[\Omega]|\Phi\rangle& = \langle\Phi| \hat{O} |\Phi\rangle \left( \langle\Phi| :\hat{O}^{\dagger}:[\Omega]|\Phi\rangle +\langle\Phi|:\hat{O}:[\Omega]|\Phi\rangle\right)~~,\nonumber\\
    \langle\Phi| \hat{H}^{(2)}[\Omega]|\Phi\rangle&=\langle\Phi| :\hat{O}^{\dagger}:[\Omega]|\Phi\rangle \langle\Phi|:\hat{O}:[\Omega]|\Phi\rangle~~,\label{vacuum}
\end{align}
where the operator $[\Omega]$ defined as $$[\Omega]=\frac{\hat{R}(\Omega)} {\langle \Phi|\hat{R}(\Omega)|\Phi\rangle}~~.$$
The rotation operator $\hat{R}(\Omega)$ is defined in Eq.~(\ref{rotop}) and the operator $\hat{O}^\dagger \hat{O}$ can be of any one of the form  $\hat{O}_n^\dagger \hat{O}_n$,  $\hat{O}_p^\dagger \hat{O}_p$,  $\hat{O}_p^\dagger \hat{O}_n$ or  $\hat{O}_n^\dagger \hat{O}_p$. In the following, it is
shown that the basic matrix element between neutron and proton quasiparticle excitations vanish. All higher-order matrix
elements between the two excited configurations can be expressed in terms of this basic matrix elements and, therefore, all them vanish. 
{\allowdisplaybreaks
\begin{align}
  \langle\Phi|a_{n_{2}^\prime} a_{n_{1}} \hat{H}[\Omega]  a^\dagger_{p_{3}} a^\dagger_{p_4^\prime} |\Phi \rangle=&
  \langle\Phi| a_{n_{2}^\prime} a_{n_{1}} \hat{H}^{(0)}[\Omega] a^\dagger_{p_{3}} a^\dagger_{p_4^\prime} |\Phi \rangle+\langle\Phi| a_{n_{2}^\prime} a_{n_{1}} \hat{H}^{(1)}[\Omega] a^\dagger_{p_{3}} a^\dagger_{p_4^\prime} |\Phi \rangle\nonumber\\
  &+\langle\Phi| a_{n_{2}^\prime} a_{n_{1}} \hat{H}^{(2)}[\Omega] a^\dagger_{p_{3}} a^\dagger_{p_4^\prime} |\Phi \rangle~~,\label{c0}
\end{align}
}
where
{\allowdisplaybreaks
\begin{align}
  \langle\Phi| a_{n_{2}^\prime} a_{n_{1}} \hat{H}^{(0)}[\Omega] a^\dagger_{p_{3}} a^\dagger_{p_4^\prime} |\Phi \rangle&={\langle\Phi| \hat{O}|\Phi \rangle}^{2} \langle\Phi| a_{n_{2}^\prime} a_{n_{1}}[\Omega] a^\dagger_{p_{3}} a^\dagger_{p_4^\prime} |\Phi \rangle\nonumber\\
  &={\langle\Phi| \hat{O}|\Phi \rangle}^{2} \left[\langle\Phi| a_{n_{2}^\prime} a_{n_{1}}[\Omega] |\Phi \rangle \langle\Phi|[\Omega] a^\dagger_{p_{3}} a^\dagger_{p_4^\prime} |\Phi \rangle-\langle\Phi| a_{n_{2}^\prime}[\Omega] a^\dagger_{p_{3}} |\Phi \rangle\langle\Phi|a_{n_{1}}[\Omega] a^\dagger_{p_4^\prime} |\Phi \rangle\right.\nonumber\\
    &\left.\hspace{2.3cm}+\langle\Phi| a_{n_{2}^\prime}[\Omega] a^\dagger_{p_4^\prime} |\Phi \rangle\langle\Phi| a_{n_{1}}[\Omega] a^\dagger_{p_{3}} |\Phi \rangle\right]\nonumber\\&=0~~.\label{c1}
\end{align}
}
In above, the terms of the type $\langle\Phi| a_{n_{2}^\prime}[\Omega] a^\dagger_{p_4^\prime} |\Phi \rangle$, $\langle\Phi| a_{n_{1}}[\Omega] a^\dagger_{p_{3}} |\Phi \rangle$ and $\langle\Phi| a_{n_{2}^\prime}[\Omega] a^\dagger_{p_{3}} |\Phi \rangle$ vanish
since $|\Phi\rangle$ is a product of neutron and proton vacuum states, i.e,$$|\Phi\rangle=|\Phi_n\rangle|\Phi_p\rangle~~,$$ and $$\langle\Phi| a_{n_{2}^\prime}[\Omega] a^\dagger_{p_4^\prime} |\Phi \rangle=\langle\Phi_n| a_{n_{2}^\prime}[\Omega]|\Phi_n\rangle\langle\Phi_p|[\Omega] a^\dagger_{p_4^\prime} |\Phi_p \rangle~~,$$ since  $|\Phi_n\rangle$ and  $|\Phi_p\rangle$ have $+$ive parity, both the overlaps on the right-hand side vanish due to parity symmetry.

Therefore,
{\allowdisplaybreaks
  \begin{align}
    \langle\Phi| a_{n_{2}^\prime}[\Omega] a^\dagger_{p_4^\prime} |\Phi \rangle=0~~.\label{overlap}
  \end{align}
  }
The second term of Eq.~(\ref{c0}) is
{\allowdisplaybreaks
  \begin{align}
    \langle\Phi|a_{n_{2}^\prime} a_{n_{1}} \hat{H}^{(1)}&[\Omega] a^\dagger_{p_{3}} a^\dagger_{p_4^\prime} |\Phi \rangle= \langle\Phi| \hat{O}|\Phi \rangle \langle\Phi|a_{n_{2}^\prime} a_{n_{1}} \left(:\hat{O}^\dagger:+:\hat{O}:\right)[\Omega] a^\dagger_{p_{3}} a^\dagger_{p_4^\prime} |\Phi \rangle\nonumber\\
     &= \langle\Phi| \hat{O}|\Phi \rangle \left[ \langle\Phi| a_{n_{2}^\prime} a_{n_{1}}:\hat{O}^{\dagger}:[\Omega]|\Phi\rangle\langle\Phi|[\Omega] a^\dagger_{p_{3}} a^\dagger_{p_4^\prime} |\Phi \rangle
    +\langle\Phi| a_{n_{2}^\prime} a_{n_{1}}:\hat{O}:[\Omega] |\Phi\rangle
 \langle\Phi| [\Omega] a^\dagger_{p_{3}} a^\dagger_{p_4^\prime}  |\Phi\rangle\right.\nonumber\\
  &\left. \hspace{0.8cm} +\langle\Phi| a_{n_{2}^\prime} a_{n_{1}}[\Omega]|\Phi\rangle\langle\Phi|:\hat{O}^{\dagger}:[\Omega] a^\dagger_{p_{3}} a^\dagger_{p_4^\prime}|\Phi \rangle
    +\langle\Phi| a_{n_{2}^\prime} a_{n_{1}}[\Omega]|\Phi\rangle\langle\Phi|:\hat{O}:[\Omega] a^\dagger_{p_{3}} a^\dagger_{p_4^\prime}|\Phi\rangle\right.\nonumber\\
  &\left. \hspace{0.8cm}  -\langle\Phi|a_{n_{2}^\prime}:\hat{O}^{\dagger}:[\Omega] a^\dagger_{p_{3}}|\Phi\rangle
    \langle\Phi|a_{n_{1}}[\Omega]a^\dagger_{p_4^\prime}|\Phi\rangle
   -\langle \Phi| a_{n_{2}^\prime}:\hat{O}:[\Omega] a^\dagger_{p_{3}}|\Phi\rangle\langle\Phi|a_{n_{1}}[\Omega]a^\dagger_{p_4^\prime}|\Phi\rangle\right.\nonumber\\
  &\left. \hspace{0.8cm} -\langle\Phi|a_{n_{2}^\prime}[\Omega] a^\dagger_{p_{3}}|\Phi\rangle
   \langle\Phi| a_{n_{1}}:\hat{O}^{\dagger}:[\Omega]a^\dagger_{p_4^\prime}|\Phi\rangle
  -\langle\Phi|a_{n_{2}^\prime}[\Omega] a^\dagger_{p_{3}}|\Phi\rangle\langle\Phi| a_{n_{1}}:\hat{O}: [\Omega]a^\dagger_{p_4^\prime}|\Phi\rangle\right.\nonumber\\
   &\left. \hspace{0.8cm}  +\langle\Phi|a_{n_{2}^\prime} :\hat{O}^{\dagger}:[\Omega]a^\dagger_{p_4^\prime}|\Phi\rangle\langle\Phi| a_{n_{1}}[\Omega] a^\dagger_{p_{3}}|\Phi\rangle
     +\langle\Phi| a_{n_{2}^\prime} :\hat{O}:[\Omega]a^\dagger_{p_4^\prime}|\Phi\rangle
     \langle\Phi| a_{n_{1}}[\Omega] a^\dagger_{p_{3}}|\Phi\rangle\right.\nonumber\\
  &\left.\hspace{0.8cm} +\langle\Phi| a_{n_{2}^\prime}[\Omega]a^\dagger_{p_4^\prime}|\Phi\rangle\langle\Phi|a_{n_{1}} :\hat{O}^{\dagger}: [\Omega] a^\dagger_{p_{3}}|\Phi\rangle
  + \langle\Phi|a_{n_{2}^\prime}[\Omega]a^\dagger_{p_4^\prime}|\Phi\rangle
  \langle\Phi| a_{n_{1}} :\hat{O}: [\Omega] a^\dagger_{p_{3}}|\Phi\rangle\right.\nonumber\\
 &\left.\hspace{0.8cm}+\left(\langle \Phi|:\hat{O}^{\dagger}: [\Omega]|\Phi\rangle +\langle \Phi|:\hat{O}: [\Omega]|\Phi\rangle\right)\langle\Phi| a_{n_{2}^\prime} a_{n_{1}}[\Omega] a^\dagger_{p_{3}} a^\dagger_{p_4^\prime} |\Phi \rangle\right]\nonumber\\&=0~~.\label{c2}
    \end{align}
}
All the above terms vanish due to Eqs.~(\ref{c1}) and (\ref{overlap}).
The third term of Eq.~(\ref{c0}) is given as
{\allowdisplaybreaks
  \begin{align}
    \langle\Phi|a_{n_{2}^\prime} a_{n_{1}} \hat{H}^{(2)}&[\Omega] a^\dagger_{p_{3}} a^\dagger_{p_4^\prime} |\Phi \rangle=\langle\Phi|a_{n_{2}^\prime} a_{n_{1}} \left(:\hat{O}^\dagger::\hat{O}:\right)[\Omega] a^\dagger_{p_{3}} a^\dagger_{p_4^\prime} |\Phi \rangle\nonumber\\
    &= \langle\Phi|a_{n_{2}^\prime} a_{n_{1}} :\hat{O}^{\dagger}: [\Omega]|\Phi \rangle \langle\Phi|:\hat{O}: [\Omega] a^\dagger_{p_{3}} a^\dagger_{p_4^\prime} |\Phi \rangle+\langle\Phi|a_{n_{2}^\prime} a_{n_{1}} :\hat{O}: [\Omega]|\Phi \rangle \langle\Phi|:\hat{O}^\dagger: [\Omega] a^\dagger_{p_{3}} a^\dagger_{p_4^\prime} |\Phi \rangle\nonumber\\
    &\hspace{0.3cm}- \langle\Phi|a_{n_{2}^\prime} :\hat{O}^{\dagger}: [\Omega] a^\dagger_{p_{3}}|\Phi \rangle \langle\Phi a_{n_{1}} :\hat{O}: [\Omega]a^\dagger_{p_4^\prime} |\Phi \rangle- \langle\Phi|a_{n_{2}^\prime} :\hat{O}: [\Omega] a^\dagger_{p_{3}}|\Phi \rangle \langle\Phi a_{n_{1}} :\hat{O}^\dagger: [\Omega]a^\dagger_{p_4^\prime} |\Phi \rangle\nonumber\\
    &\hspace{0.3cm}+\langle\Phi|a_{n_{2}^\prime} :\hat{O}^{\dagger}: [\Omega]a^\dagger_{p_4^\prime} |\Phi \rangle \langle\Phi| a_{n_{1}} :\hat{O}: [\Omega] a^\dagger_{p_{3}}|\Phi \rangle+\langle\Phi|a_{n_{2}^\prime} :\hat{O}: [\Omega]a^\dagger_{p_4^\prime} |\Phi \rangle \langle\Phi| a_{n_{1}} :\hat{O}^\dagger: [\Omega] a^\dagger_{p_{3}}|\Phi \rangle\nonumber\\
    &+ \langle\Phi| :\hat{O}^{\dagger}: [\Omega]|\Phi \rangle\left[\langle\Phi|a_{n_{2}^\prime} a_{n_{1}} :\hat{O}: [\Omega]|\Phi \rangle \langle\Phi|[\Omega] a^\dagger_{p_{3}} a^\dagger_{p_4^\prime} |\Phi \rangle+\langle\Phi|a_{n_{2}^\prime} a_{n_{1}} [\Omega]|\Phi \rangle \langle\Phi|:\hat{O}:[\Omega] a^\dagger_{p_{3}} a^\dagger_{p_4^\prime} |\Phi \rangle\right.\nonumber\\
      &\left.\hspace{1cm}- \langle\Phi|a_{n_{2}^\prime} :\hat{O}: [\Omega] a^\dagger_{p_{3}}|\Phi \rangle \langle\Phi a_{n_{1}} [\Omega]a^\dagger_{p_4^\prime} |\Phi \rangle-\langle\Phi|a_{n_{2}^\prime} [\Omega] a^\dagger_{p_{3}}|\Phi \rangle \langle\Phi a_{n_{1}} :\hat{O}: [\Omega]a^\dagger_{p_4^\prime} |\Phi \rangle\right.\nonumber\\
      &\left.\hspace{1cm}+\langle\Phi|a_{n_{2}^\prime} :\hat{O}: [\Omega]a^\dagger_{p_4^\prime} |\Phi \rangle \langle\Phi| a_{n_{1}} [\Omega] a^\dagger_{p_{3}}|\Phi \rangle+\langle\Phi|a_{n_{2}^\prime}[\Omega]a^\dagger_{p_4^\prime} |\Phi \rangle \langle\Phi| a_{n_{1}} :\hat{O}: [\Omega] a^\dagger_{p_{3}}|\Phi \rangle\right]\nonumber\\
      &+ \langle\Phi| :\hat{O}: [\Omega]|\Phi \rangle\left[\langle\Phi|a_{n_{2}^\prime} a_{n_{1}} :\hat{O}^\dagger: [\Omega]|\Phi \rangle \langle\Phi|[\Omega] a^\dagger_{p_{3}} a^\dagger_{p_4^\prime} |\Phi \rangle+\langle\Phi|a_{n_{2}^\prime} a_{n_{1}} [\Omega]|\Phi \rangle \langle\Phi|:\hat{O}^\dagger:[\Omega] a^\dagger_{p_{3}} a^\dagger_{p_4^\prime} |\Phi \rangle\right.\nonumber\\
      &\left.\hspace{1cm}- \langle\Phi|a_{n_{2}^\prime} :\hat{O}^\dagger: [\Omega] a^\dagger_{p_{3}}|\Phi \rangle \langle\Phi a_{n_{1}} [\Omega]a^\dagger_{p_4^\prime} |\Phi \rangle-\langle\Phi|a_{n_{2}^\prime} [\Omega] a^\dagger_{p_{3}}|\Phi \rangle \langle\Phi a_{n_{1}} :\hat{O}^\dagger: [\Omega]a^\dagger_{p_4^\prime} |\Phi \rangle\right.\nonumber\\
      &\left.\hspace{1cm}+\langle\Phi|a_{n_{2}^\prime} :\hat{O}^\dagger: [\Omega]a^\dagger_{p_4^\prime} |\Phi \rangle \langle\Phi| a_{n_{1}} [\Omega] a^\dagger_{p_{3}}|\Phi \rangle+\langle\Phi|a_{n_{2}^\prime}[\Omega]a^\dagger_{p_4^\prime} |\Phi \rangle \langle\Phi| a_{n_{1}} :\hat{O}^\dagger: [\Omega] a^\dagger_{p_{3}}|\Phi \rangle\right]\nonumber\\
    &\hspace{1cm}+\left[\langle\Phi|:\hat{O}^\dagger: [\Omega]|\Phi \rangle\langle\Phi|:\hat{O}: [\Omega]|\Phi \rangle\right]\langle\Phi| a_{n_{2}^\prime} a_{n_{1}}[\Omega] a^\dagger_{p_{3}} a^\dagger_{p_4^\prime} |\Phi \rangle\nonumber\\&=0~~.\label{c3}
    \end{align}
}
Again all the above terms vanish due to Eqs.~(\ref{c1}) and (\ref{overlap}).
Therefore, we have $$\langle\Phi|a_{n_{2}^\prime} a_{n_{1}} \hat{H}[\Omega] a^\dagger_{p_{3}} a^\dagger_{p_4^\prime} |\Phi \rangle=0~~.$$
\twocolumngrid
\bibliographystyle{apsrev4-2}
\bibliography{jacwit}

\end{document}